\documentclass[usenatbib]{mnras}

\usepackage{newtxtext,newtxmath}

\usepackage[T1]{fontenc}
\usepackage{ae,aecompl}
\usepackage{tabularx}

\usepackage{graphicx}	
\usepackage{amsmath}	
\usepackage{amssymb}	

\newcommand{\tfifty}{$t_{50}$ }
\newcommand{\tninty}{$t_{90}$ }

\newcommand{\msun}{{\rm M}_\odot}
\newcommand{\vmax}{$V_{\rm max}$ }

\title[SFH gradients]{A Predicted Correlation Between Age Gradient and Star Formation History in FIRE Dwarf Galaxies}
\author[A. S. Graus et al.]{Andrew S. Graus,$^{1,2}$\thanks{E-mail: agraus@utexas.edu}
James S. Bullock,$^{2}$
Alex Fitts,$^{1}$
Michael C. Cooper,$^{2}$ \newauthor
Michael Boylan-Kolchin,$^{1}$
Daniel R. Weisz,$^{3}$
Andrew Wetzel,$^{4}$  \newauthor
Robert Feldmann,$^{5}$
Claude-Andr\'e Faucher-Gigu\`ere$^{6}$
Eliot Quataert$^{3}$ \newauthor
Philip F. Hopkins$^{7}$
Du\u san Kere\u s$^{8}$
\\
$^{1}$Department of Astronomy, The University of Texas at Austin, 2515 Speedway Stop C1400, Austin, TX 78712, USA \\
$^{2}$Center for Cosmology, Department of Physics and Astronomy,4129 Reines Hall, University of California Irvine, CA 92697, USA \\
$^{3}$Department of Astronomy and Theoretical Astrophysics Center, University of California, Berkeley, Berkeley, CA 94720, USA\\
$^{4}$Department of Physics, University of California, Davis, CA 95717, USA\\
$^{5}$Institute for Computational Science, University of Zurich, Zurich CH-8057, Switzerland\\
$^{6}$Department of Physics and Astronomy and CIERA, Northwestern University, 2145 Sheridan Road, Evanston, IL 60208, USA\\
$^{7}$TAPIR, Mailcode 350-17, California Institute of Technology, Pasadena, CA 91125, USA\\
$^{8}$Department of Physics, Center for Astrophysics and Space Science, University of California at San Diego, 9500 Gilman Drive, \\ La Jolla, CA 92093, USA
}

\date{Accepted XXX. Received YYY; in original form ZZZ}

\pubyear{2018}

\begin{document}
\label{firstpage}
\pagerange{\pageref{firstpage}--\pageref{lastpage}}
\maketitle

\begin{abstract}
We explore the radial variation of star formation histories in dwarf galaxies simulated with Feedback In Realistic Environments (FIRE) physics. The sample contains 9 low-mass field dwarfs with $M_{\rm star} = 10^5 - 10^7 \msun$ from previous FIRE results, and a new suite of 17 higher mass field dwarfs with $M_{\rm star} = 10^7 - 10^9 \msun$ introduced here. We find that age gradients are common in our dwarfs, with older stars dominant at large radii. The strength of the gradient correlates with overall galaxy age such that earlier star formation produces a more pronounced gradient. The relation between formation time and strength of the gradient is driven by both mergers and star-formation feedback. Mergers can both steepen and flatten the age gradient depending on the timing of the merger and star formation history of the merging galaxy. In galaxies without significant mergers, early feedback pushes stars to the outskirts at early times.  Interestingly, among galaxies without mergers, those with large dark matter cores have flatter age gradients because these galaxies have more late-time feedback.  If real galaxies have age gradients as we predict, stellar population studies that rely on sampling a limited fraction of a galaxy can give a biased view of its global star formation history. We show that central fields can be biased young by a few Gyrs while outer fields are biased old. Fields positioned near the 2D half-light radius will provide the least biased measure of a dwarf galaxy's global star formation history. 
\end{abstract}




\begin{table*}
  \caption{Table of properties of the simulated galaxies used in this work at $\it{z}$ = 0. (1) stellar mass,  (2) halo mass, (3) maximum circular velocity, (4) mean 2D half-light radius over all projections; (5) lookback time to the formation of 50\% of stars within 10\% of the virial radius of the halo; (6) lookback time to the formation of 90\% of stars within 10\% of the virial radius of the halo; (7) and (8) age gradients defined in Equation \ref{eq:gamma_eq}.  }
\centering 
\begin{tabularx}{\textwidth}{lXXXXXXXX}
\hline
\hline  
  &       $M_{\rm star}$ &        $M_{\rm halo}$ &      $\rm V_{max}$ &    $\rm R_{1/2}$ & \tfifty & \tninty & $\gamma_{50}$ & $\gamma_{90}$\\
  &  $\rm [\msun]$ & $\rm [\msun]$ & [$\rm kms^{-1}$] & [kpc] & [Gyr] & [Gyr] &[Gyr/$\rm R_{1/2}$] & [Gyr/$\rm R_{1/2}$]\\
\hline 
Halo & (1) & (2) & (3) & (4) & (5) & (6) & (7) & (8)\\
\hline

m10xa &   7.64e07 &  1.87e10 &  45.26 &  2.23 & 6.08 & 1.06 & -0.3 & 0.1 \\
m10xb &   3.29e07 &  2.22e10 &  42.78 &  1.73 & 4.23 & 0.85 & -0.5 & -0.6 \\
m10xc &  1.19e08 &  3.22e10 &  48.31 &  2.25 & 6.55 & 1.00 & -2.2 & -0.3 \\
m10xc\_A &  8.46e06 &  8.52e09 &  35.03 &  1.24 & 10.89 & 4.55 & -2.4 & -4.0 \\
m10xd &   6.81e07 &  3.86e10 &  53.51 &  2.60 & 4.04 & 1.72 & -0.5 & -1.0 \\
m10xd\_A &   1.44e07 &  2.40e10 &  38.52 &  1.38 & 1.63 & 0.469 & 0.0 & -0.1 \\
m10xe &  3.26e08 &  4.57e10 &  56.17 &  2.93 & 6.13 & 1.72 & -1.5 & -0.9\\
m10xe\_A &    3.64e06 &  1.36e10 &  35.74 &  0.90 & 8.50 & 1.17 & -3.2 & -1.4 \\
m10xe\_B &   1.28e07 &  1.12e10 &  38.15 &  1.31 & 8.75 & 4.76 & -1.8 & -2.9 \\
m10xe\_C &   1.84e07 &  1.04e10 &  34.43 &  2.11 & 7.08 & 1.66 & -0.9 & -0.6 \\
m10xe\_D &    3.61e06 &  8.88e09 &  34.13 & 2.43 & 9.62 & 3.82 & 0.1 & -0.9\\
m10xf &  1.28e08 &  5.21e10 &  58.47 &  2.30 & 7.38 & 1.93 & -2.6 & -2.6 \\
m10xg &  4.61e08 &  6.20e10 &  65.75 &  2.78 & 7.59 & 2.00 & -2.3 & -1.3 \\
m10xg\_A &   1.88e07 &  1.53e10 &  40.31 &  1.51 & 5.11 & 0.86 & -1.3 & -0.3 \\
m10xh &  5.4e08 &  7.44e10 &  68.10 &  4.15 & 3.65 & 0.55 & 0.9 & 0.2 \\
m10xh\_A &   4.97e07 &  1.47e10 &  38.80 &  2.19 & 5.68 & 1.39 & -1.6 & -0.7 \\
m10xi &  4.48e08 &  7.58e10 &  64.35 &  3.56 & 6.03 & 2.71 & -2.1 & -1.6\\
\hline

\citet{Fitts17} &  &  &  & \\

\hline
m10b &     4.65e+05 &  9.29e+09 &  31.51 &  0.24 & 2.54 & 0.65 & -0.1 & -0.1 \\
m10c &	   5.75e+05 &  8.92e+09 &  31.40 &  0.25 & 4.07 & 0.96 & -1.8 & -0.6 \\
m10e &	   1.98e+06 &  1.02e+10 &  31.44&  0.43 & 5.63 & 1.02 & -0.9 & -0.6 \\
m10f &     4.11e+06 &  8.56e+09 &  35.66 &  0.52 & 11.96 & 5.33 & -4.0 & -4.2 \\
m10h &     7.80e+06 &  1.28e+10 &  37.98 &  0.58 & 11.64 & 2.52 & -4.6 & -2.5 \\
m10j &     9.74e+06 &  1.10e+10 &  37.98 &  0.50 & 11.51 & 3.94 & -2.5 & -2.7 \\
m10k &     1.04e+07 &  1.15e+10 &  38.22 &  0.85 & 10.74 & 4.18 & -2.5 & -2.5 \\
m10l &     1.30e+07 &  1.06e+10 &  37.62 &  0.54 & 10.76 & 3.34 & -2.1 & -3.0 \\
m10m &     1.44e+07 &  1.15e+10 &  38.51&  0.69 & 9.86 & 3.76 & -2.1 & -3.1 \\
\hline
\label{table:galaxy_stats}
\end{tabularx}
\end{table*}

\section{Introduction} \label{s:intro}

A key question in galaxy formation is to understand how stellar mass builds up in galaxies over time.   Observed color-magnitude diagrams (CMDs) together with sophisticated stellar population synthesis models provide a  powerful approach to measure the star formation histories (SFHs) of  galaxies and directly answer this question for certain systems \citep{Dolphin02,Tolstoy09,Monelli10,Walmswell13,Cole14,Weisz14,Monelli16,Makarova17,Skillman17}. This technique is particularly useful for understanding dwarf galaxies in the local Universe, where precise photometry for populations of individual stars enables the construction of accurate CMDs.

The inferred SFHs of local dwarf galaxies have revealed much about the nature of galaxy formation on small scales.  For example, ultrafaint dwarfs appear to be almost universally old \citep{Brown14,Weisz15}, lending support to the idea that these objects had their star formation quenched by reionization \citep{Efstathiou92,Bullock00,Ricotti05,Wimberly18}.  Most larger dwarf galaxies, on the other hand, cease star formation only when they are within the virial radius of a larger galaxy (e.g. the Milky Way or M31), a result that provides a useful means to constrain models of environmental quenching \citep{Geha12,Gallart15,Weisz15,Wetzel15,Fillingham15,Fillingham16}. 

A further application of Local Group dwarf SFHs is to study the high-redshift universe. Accurate measurements of SFHs allow us to extrapolate galaxy properties back in time and to place constraints on high-redshift luminosity and stellar mass functions \citep{MBK2014,Weisz14b,MBK2015,Graus16}.  More generally, comparisons between SFHs from simulated and observed galaxies provide important tests for cosmological models of galaxy formation. Such studies suggest that strong stellar feedback is essential for explaining the dwarf galaxy population \citep[e.g.,][]{DiCintio14,Dutton16,Sawala16b,Read16,SGK17}.

Strong stellar feedback not only regulates star formation, it also can change structural properties of the galaxy. For example, feedback from supernovae has been shown to create cores in the dark matter halo profiles of simulated dwarf galaxies \citep[e.g.,][]{Pontzen12,Onorbe15,Fitts17}. This is important as it could be a key component in solving small-scale problems with $\Lambda$CDM \citep{BBK17}. For example, the cusp-core problem \citep{Flores94,Moore94}, where observations of some dwarf galaxies are best fit by cored dark matter density profiles, in contrast to dark matter only simulations in which halos have NFW cusps. This can also help alleviate  the ``Too Big To Fail" problem \citep{MBK2011}, where the most massive subhalos of Milky Way-like halos appear too dense to host the largest galaxies seen observationally in $\Lambda$CDM.

The same feedback episodes that alter the dynamics of dissipationless dark matter in halos can also affect the collisionless stars in galaxies. Indeed, we expect the dark matter and stars to respond dynamically to feedback-induced potential fluctuations in a qualitatively similar manner, given that both behave as (effectively) collisionless fluids. Such an effect was investigated by \cite{ElBadry16}, who used the FIRE-1\footnote{\url{http://fire.northwestern.edu}} simulations to show that simulated dwarf galaxies with strong stellar feedback have large fluctuations in their effective stellar radii over time \citep[see also][]{Stinson09}. This effect can eventually lead to an overall median age gradient where young stars form in the center of the galaxy and old stars are preferentially found in the outskirts.
Qualitatively, this agrees with the observed age and metallicity gradients seen in most dwarf galaxies locally, where younger (more metal-rich) stars lie in the center and older (more metal-poor) stars in the outskirts \citep{Battaglia06,Faria07,Beccari14,McMonigal14,delPino15,Santana16,Kacharov17,McQuinn17,Okamoto17,Cicuendez18a,Cicuendez18b}.

\begin{figure*}
	\includegraphics[width=0.49\textwidth, trim = 0 0 0 0]{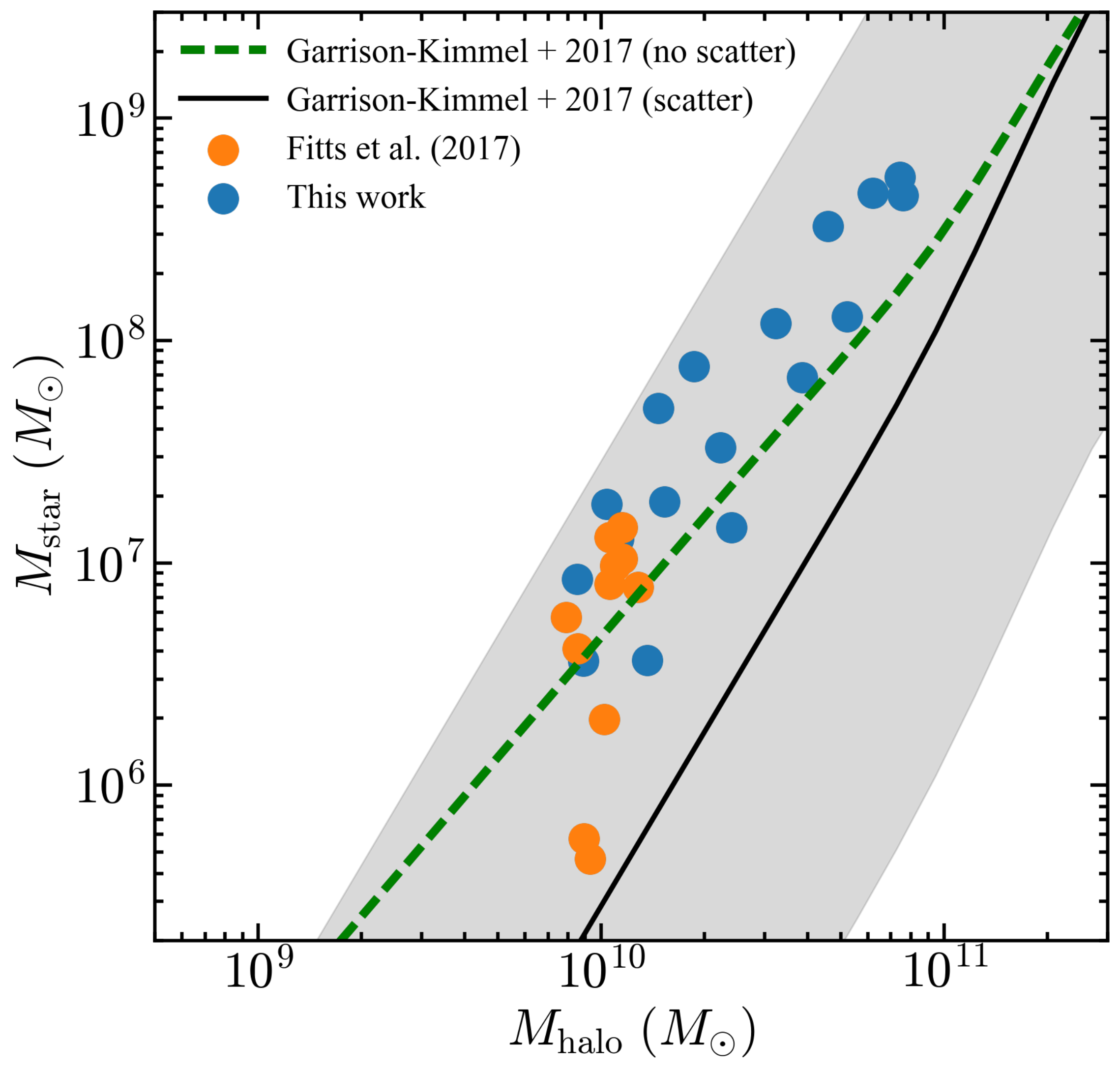}
	\includegraphics[width=0.49\textwidth, trim = 3.0cm 0.5cm 3.0cm 3.0cm]{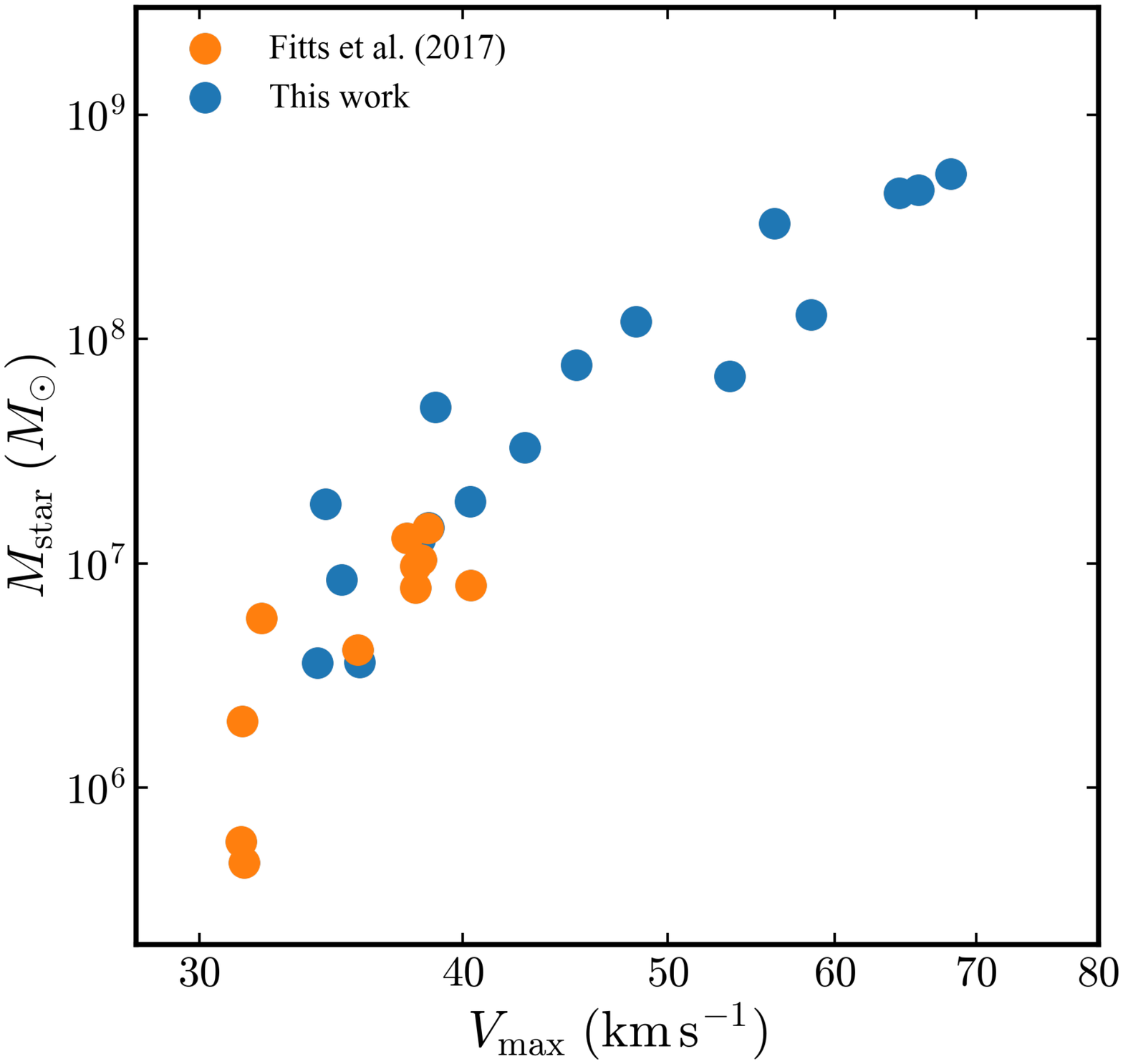}
	\centering
	\caption[]{Stellar mass vs. halo mass (left) and stellar mass vs. halo maximum circular velocity (right) for the simulated dwarf galaxies used in this paper.  The dashed and solid lines in the left panel show abundance matching relations presented in \citet{SGK17}, which show the best fit abundance matching relation for the Milky Way satellites given zero scatter (green) and 2 dex of scatter (black line with the grey band).}
	\label{fig:abundance_matching}
\end{figure*}

Age and metallicity gradients could potentially be relevant for the measurement of SFHs, as a spread in age with position could lead to biases in the observed SFHs relative to the true SFH.  Specifically, if a CMD study relies on a field that is small compared to the galaxy's area on the sky, the inferred SFH might not represent the global history of the galaxy.  While the simulations studied in \cite{ElBadry16} included enough dwarf galaxies to study the origin of population gradients (four systems with $M_{\rm star} = 10^6 - 10^9$ $\msun$) the sample was not large enough to explore trends and potential observational biases associated with this effect. Here we use a sample of 26 dwarf galaxy simulations with stellar masses from $10^5 - 10^9$ $\msun$, including 9 presented in \citet{Fitts17} and 17 introduced here, all  run with the \texttt{GIZMO} code \citep{Hopkins15}\footnote{\url{http://www.tapir.caltech.edu/~phopkins/Site/GIZMO.html}} and the FIRE-2 feedback implementation \citep{Hopkins18}.  Our aim is to study the spatial variation in SFHs over our entire suite of simulations, to search for correlations between the SFH gradient and other observables, and to use our simulations to explore the biases in SFHs that could arise from small-field CMD studies of local galaxies. 

In section \ref{s:sims and methods} we discuss the simulations and our methodology for measuring variations in SFHs.  Section \ref{s:results} presents our predictions for SFHs, how they vary with radius, and shows that the strength of the gradient increases with galaxy age. In section \ref{sec:discussion}, we discuss the origins of these variations, and what the gradients imply for interpreting current and future observations of Local Group dwarf galaxies.

\section{Simulations and Methods} \label{s:sims and methods}

The simulations used in this work were run using the multi-method gravity+hydrodynamics code \texttt{GIZMO} \citep{Hopkins15} and utilize the the FIRE-2 feedback implementation \citep{Hopkins18}.  We specifically use a mesh-free Lagrangian Godunov (MFM) method that is second-order accurate and maintains many of the advantages of traditional SPH codes, while avoiding some of the traditional pitfalls of classic SPH codes such as accurate capturing of shocks, and general treatment of fluids, for which grid-based codes have traditionally been better. The simulations include star formation in dense molecular gas that is self-shielding, and Jeans-unstable \citep{Krumholz11}. We also include cooling and heating from an ionizing background \citep{Faucher09}, along with photo-heating and radiation pressure from stellar sources including feedback from OB stars and AGB mass-loss. We also include Type Ia and Type II supernovae. All the stellar physics is calculated assuming each star particle is a simple stellar population with a \citet{Kroupa01} initial mass function. Furthermore, we include turbulent diffusion of metals \citep{Hopkins17diff,Su17}, which provides for a better match to observed metallicity distributions of Local Group dwarf galaxies \citep{Escala18}.

Our sample includes 9 of the dwarf galaxies introduced in \cite{Fitts17} along with an additional suite of 17 isolated dwarf galaxies presented for the first time here.  Table \ref{table:galaxy_stats} presents an overview of the simulations along with adopted names.   The \cite{Fitts17} dwarfs  (named m10b-m) were choose to form in halos of mass  $M_{\rm v} \simeq$ $10^{10}$ $\msun$ at $z=0$ and were simulated with dark matter particle masses of $m_{\rm dm}$ = 2500 $\msun$ and initial gas particle masses of $m_{\rm g}$ = 500 $\msun$ These dwarf galaxies form between $10^{5}$ and $10^{7}$ $\msun$ in stars, where the mass of a star particle is initially the same as a gas particle. The second set of simulations (named m10xa-i) includes 17 halos with $M_{\rm v} = 0.1-1 \times  10^{11}$ $\msun$ each run with dark matter masses of $m_{\rm dm}$ = 20000 $\msun$ and initial gas particle masses of $m_{\rm g}$ = 4000 $\msun$. These more massive dwarfs form between $10^{7}$ and $10^{9}$ $\msun$ of stars and are similar to the most massive dwarf galaxies seen in the Local Group. For the new suite of simulations we name the most massive halo after the simulation itself (m10xa - i). Lower mass halos from the same simulation are designated by the simulation name followed by a capital latter (A, B, C, ect.). The stellar mass vs. halo mass and stellar mass vs. \vmax relation for the simulated galaxies used in this work are shown in Figure \ref{fig:abundance_matching}. 

All of these simulations are cosmological zoom-in simulations \citep[e.g.,][]{Katz93,Onorbe14} with initial conditions generated using the \texttt{MUSIC} initial conditions generator \citep{Hahn11}. Halo finding in the simulation was done using a combination of the {\tt Rockstar} halo finder \citep{Behroozi13}, and the Amiga Halo Finder \citep[\tt{AHF}]{Knollmann09} to verify the properties of the halos and galaxies in this suite including the masses and centers of the halos and galaxies. One final note is that the two sets of simulations have slightly different cosmologies with the \cite{Fitts17} sample having cosmological parameters: $H_0$ = 71.0 km s$^{-1}$ Mpc$^{-1}$, $\Omega_{\rm m}$ = 0.266, $\Omega_{\rm b}$ = 0.044, $\Omega_{\Lambda}$ = 0.734, while for the new sample of large galaxies the cosmological parameters are $H_0$ = 70.2 km s$^{-1}$ Mpc$^{-1}$, $\Omega_{\rm m}$ = 0.272, $\Omega_{\rm b}$ = 0.0455, $\Omega_{\Lambda}$ = 0.728.

\section{Results} \label{s:results}
\subsection{SFH gradients}

\begin{figure*}
	\includegraphics[width=0.95\textwidth,height=0.25\textheight, trim = 0 0 0 0]{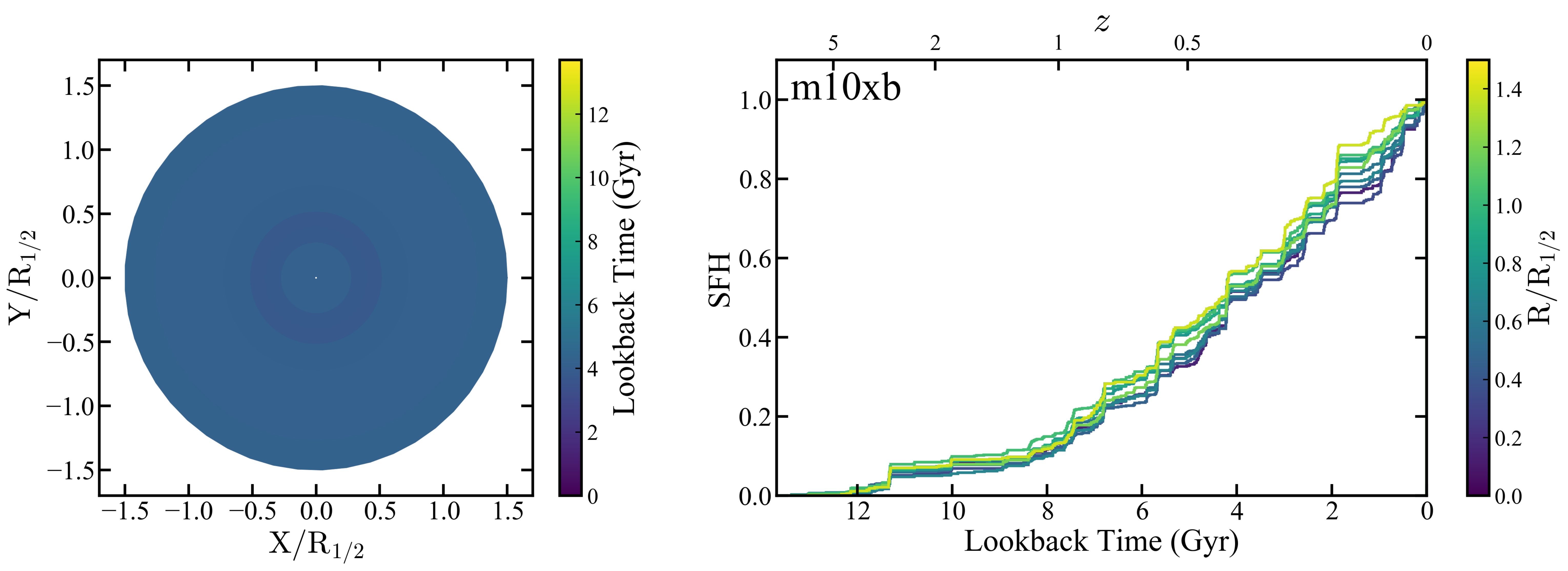}
	\hspace{.2in}
	\includegraphics[width=0.95\textwidth,height=0.25\textheight, trim = 0 0 0 0]{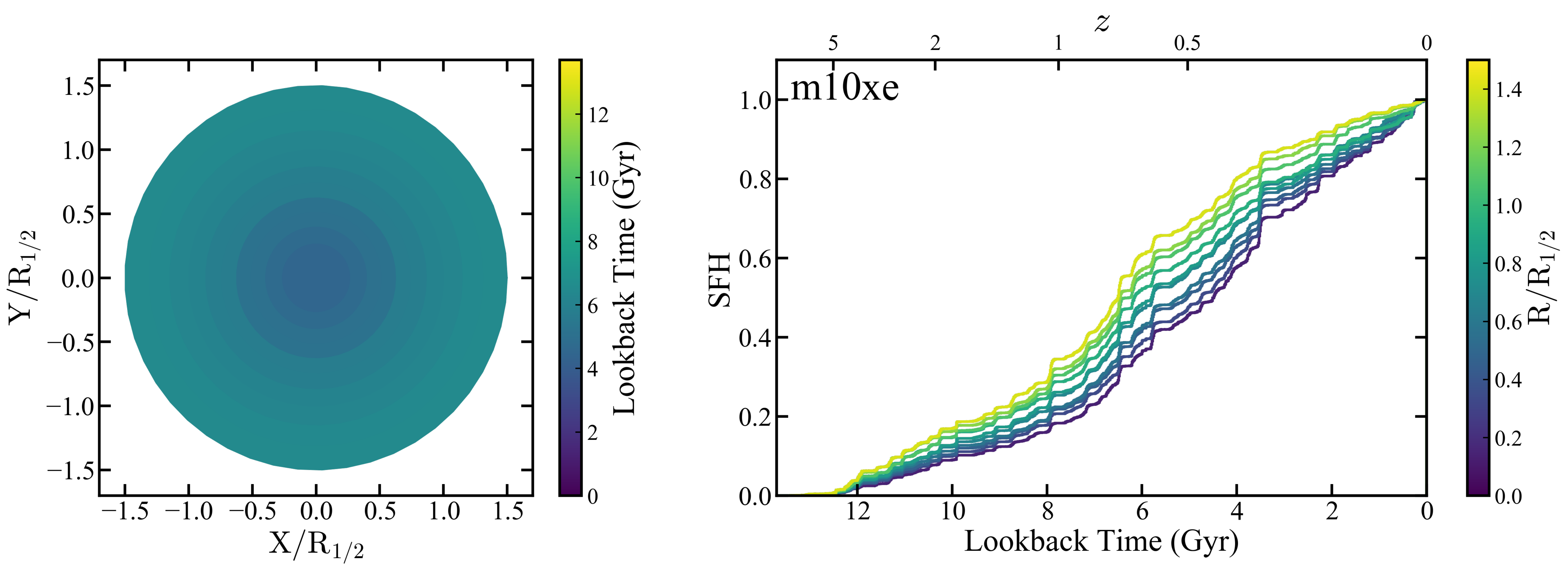}
	\hspace{.2in}
	\includegraphics[width=0.95\textwidth,height=0.25\textheight, trim = 0 0 0 0]{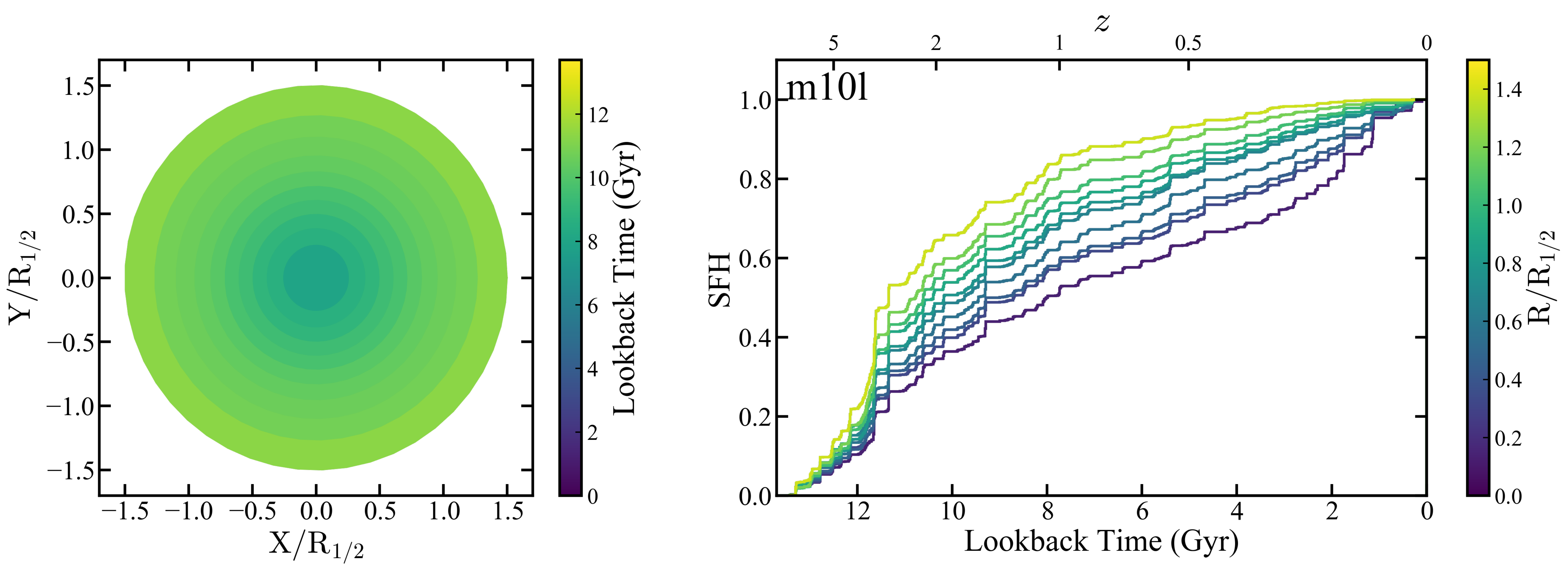}
	\centering
	\caption[]{Example star formation history gradients in three galaxies.  \textit{Left: } Map of median stellar age in projected radial bins normalized by the 2D half-mass radius. The color bar maps to age (lookback time to formation) as indicated, with yellow corresponding to older stars and purple to younger stars. \textit{Right: } The cumulative SFH within each projected radial bin colored by the distance of the bin from the center of the galaxy as indicated by the color bar on the right. Note that the youngest galaxy has the least pronounced SFH gradient on the upper right. The oldest galaxy has the strongest SFH gradient on the lower right. The bins are fixed to be circular, however we note that the resulting gradients do not change significantly if we allow the contours to be elliptical.}
	\label{fig:grad_map}
\end{figure*}

\begin{figure*}
	\includegraphics[width=0.32\textwidth, trim = 0 0 0 0]{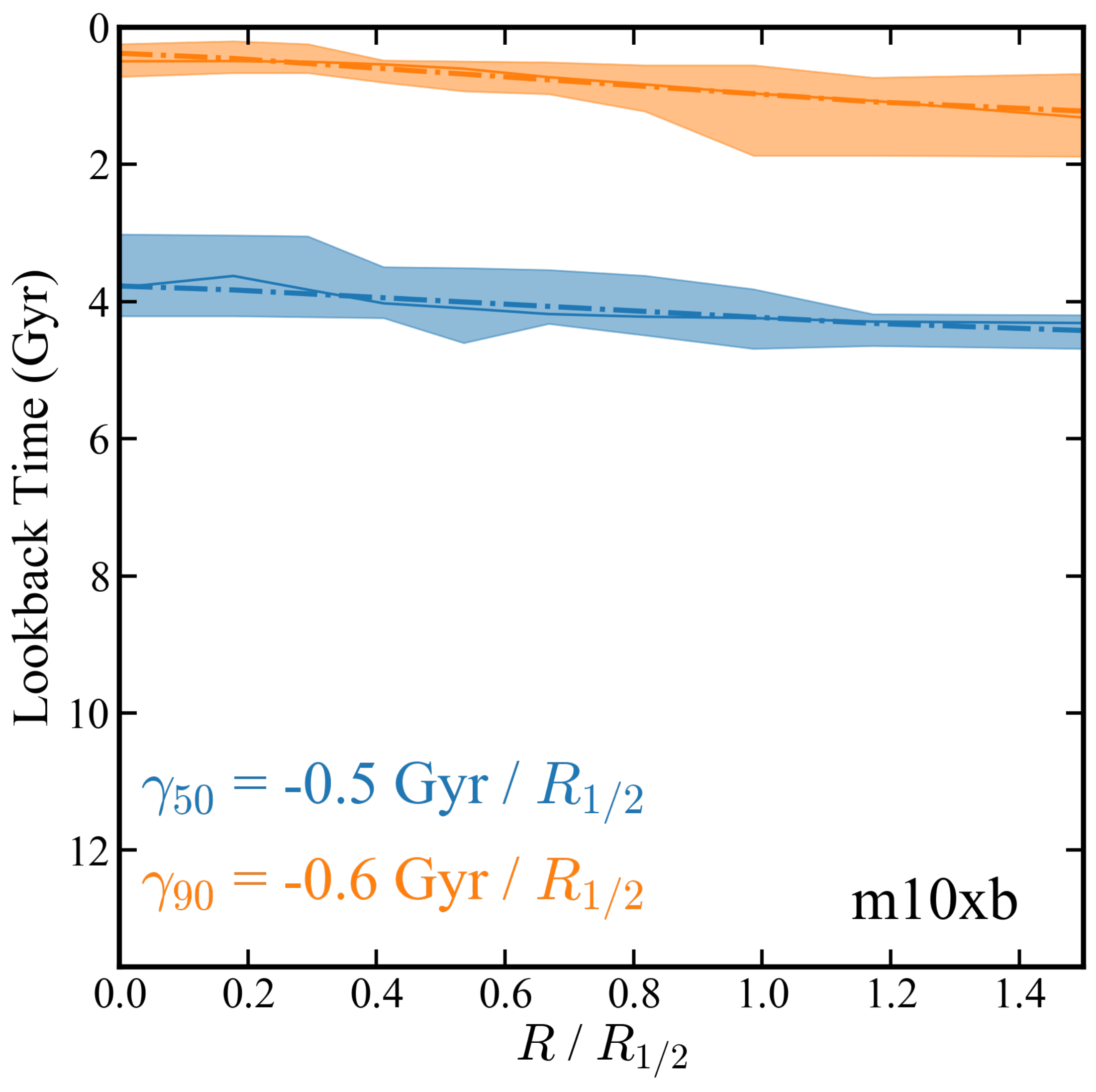}
	\includegraphics[width=0.32\textwidth, trim = 0 0 0 0]{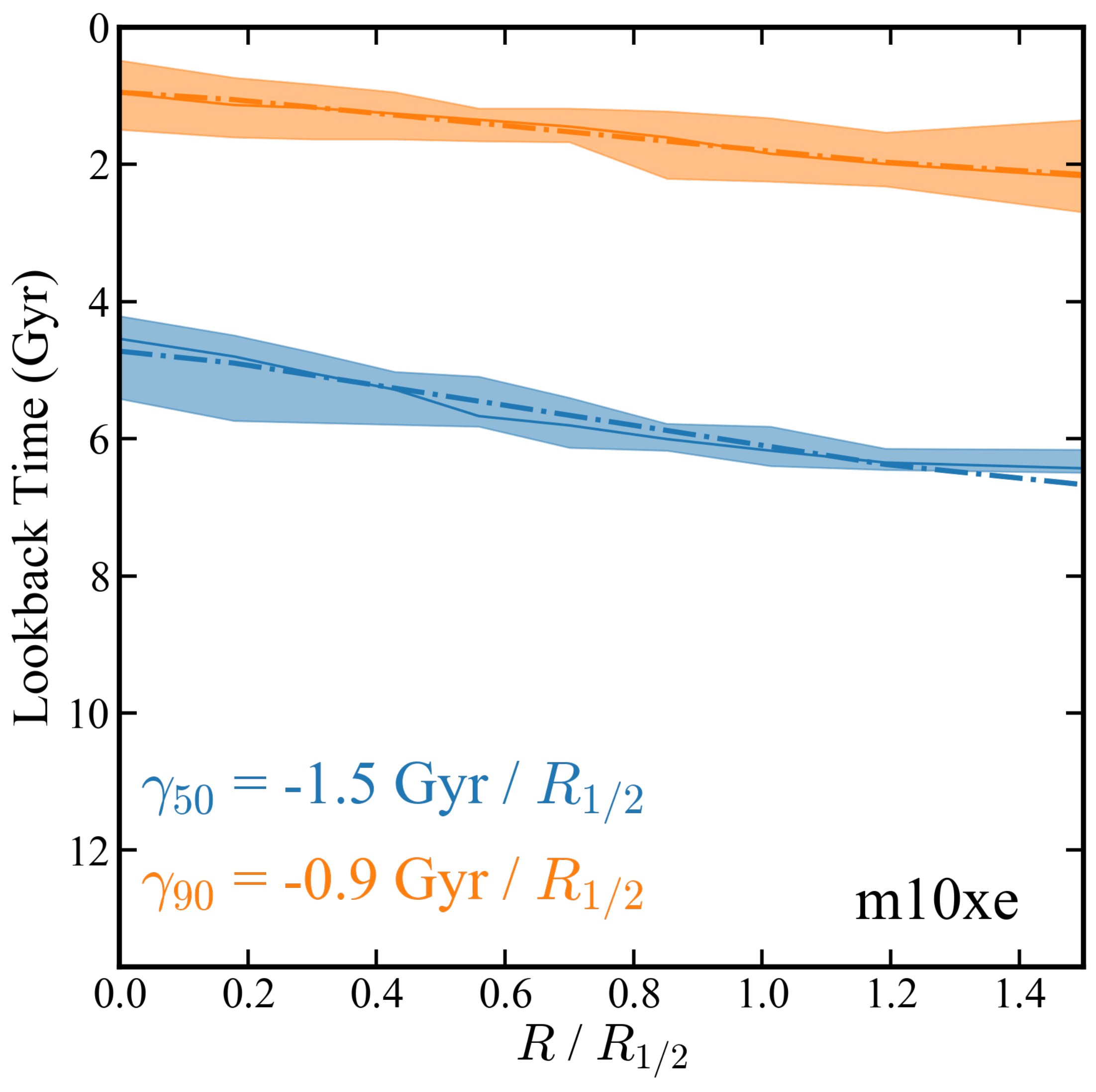}
	\includegraphics[width=0.32\textwidth, trim = 0 0 0 0]{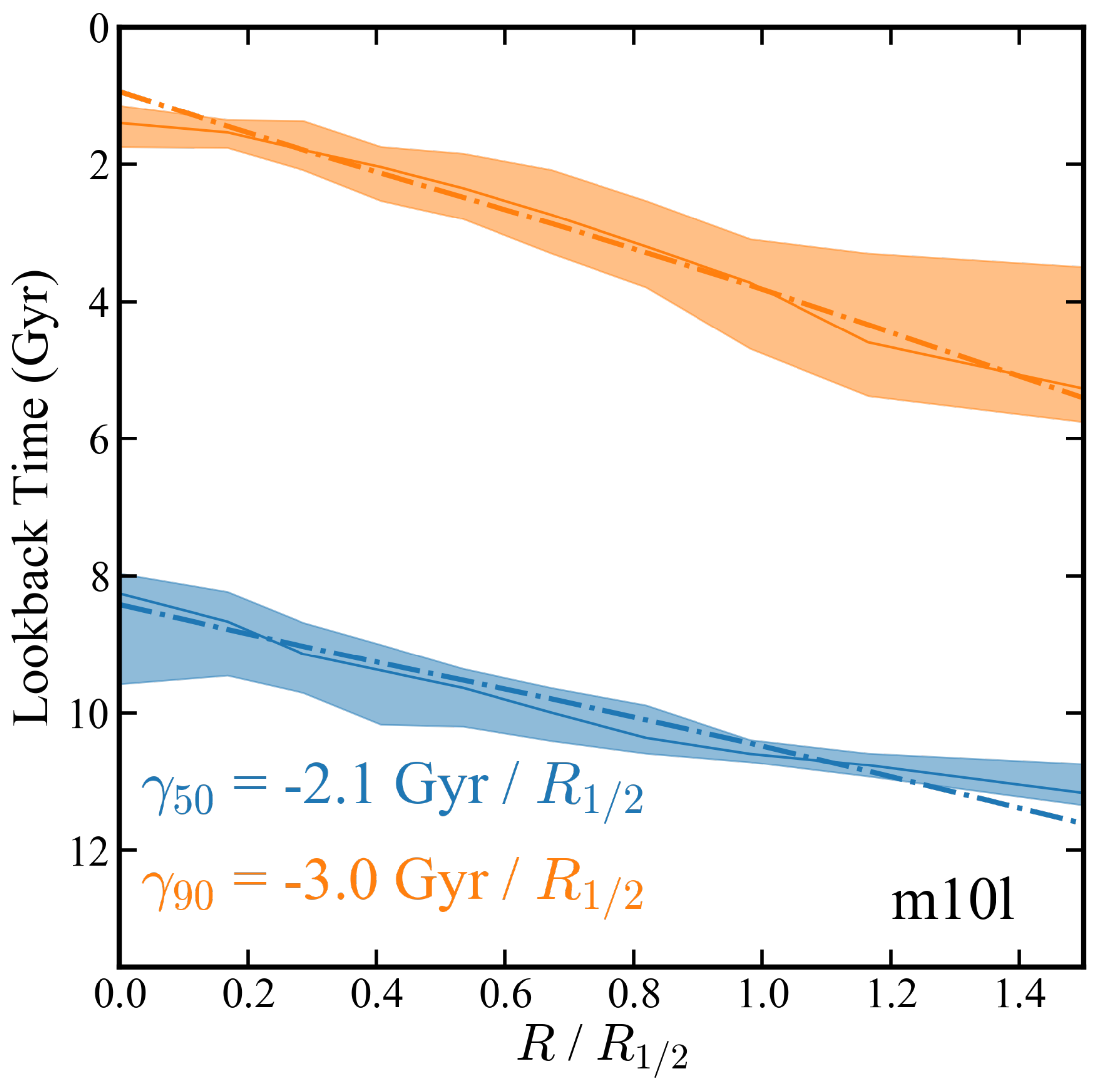}
	\centering
	\caption[]{Age vs. projected radius for three example galaxies  -- m10xb, m10xe, and m10l from left to right, respectively.  These are the same galaxies illustrated in Figure \ref{fig:grad_map}. The times shown are specifically the lookback times to 50\% star formation ($t_{50}$, blue) and the lookback times to  90\% star formation ($t_{90}$, orange). The solid line represents the median age measured at each projected radius for 100 random viewing angles for each galaxy, while the shaded bands show the 99th percentile range in measured age at a given projected radius over all 100 random viewing angles. The corresponding slope of the gradient as defined in Equation \ref{eq:gamma_eq} is shown in the lower left of each panel.}
	\label{fig:gradients}
\end{figure*}

Figure \ref{fig:grad_map} shows examples of the SFH gradients for three galaxies in the sample (m10xb, m10xe, m10l) arranged from top to bottom by strength of the gradient. The images in the left column show circularly-averaged age maps seen in projection along an arbitrary axis.  The size scale is normalized by the projected (2D) half-mass radius of the galaxy $R_{1/2}$. The stars are then binned into 10 bins within 1.5  $\times$ $R_{1/2}$ such that the number of stars is equal in each bin. The color code maps to the median age of stars in each radial bin. The panels on the right show the SFHs within ten radial bins, now color-coded by the bin radius. The top row shows one of the weakest SFH gradients in our sample, while the bottom row shows one of the strongest.  Note that m10l with a strong gradient is also the oldest of the three (with the longest lookback time for median star formation) while m10xb is the youngest overall.  We show below that a trend between global age and overall SFH gradient is seen throughout our sample.

Figure \ref{fig:gradients} quantifies the age gradients for the same galaxies illustrated in Figure \ref{fig:grad_map}. The two colors correspond to different characteristic ages: blue shows the lookback time to when stars in a projected radial bin first reached 50\% of their final stellar mass ($t_{50}$); orange shows the lookback time to the formation of 90\% of the final stellar mass ($t_{90}$).  Stated another way, \tfifty is the median age of the stars at a given radius and \tninty is the 90th percentile age of stars at a given radius.  The solid blue (orange) line shows the median values of \tfifty ($t_{90}$) computed at a given projected radius over 100 random viewing angles for each galaxy. The shaded regions show the 99th percentile ranges over all projections. We quantify the gradients by measuring the radial variation in $t_{50}$ and $t_{90}$ as:

\begin{equation}
 \gamma_{50} = \frac{\Delta t_{50}}{\Delta R / R_{1/2}} \qquad ; \qquad \gamma_{90} = \frac{\Delta t_{90}}{\Delta R / R_{1/2}}.
\label{eq:gamma_eq}
\end{equation}

Note that the gradients are normalized to the half-mass radius and have units of Gyr (per $R_{1/2}$). The value of the slope tells us simply how much older the stars at the half-mass radius are than those at the center of the galaxy in Gyr. We measure the slope of the gradient in two different ways.  First, we take the difference between $t_{50}$ and $t_{90}$ in the innermost and outermost bins, and then divide it by the difference in the mean radius of the inner and outer bins. To double check this, we also measure the slope by calculating the least squares fit to a line, and taking the slope of that line. We find that the different methods of measuring the slope make little difference, and the variation in gradient over projection angle is a much larger effect.

The median values for the gradients we measure for each galaxy in our simulated sample are listed in Table \ref{table:galaxy_stats}. We see that most of the dwarf galaxies show a clear negative age gradient, where the stars are younger in the inner regions and older in the outskirts. Some of the gradients are small, or consistent with being flat, but most are clearly negative. The gradients are measured within 1.5 $\times$ $R_{1/2}$ by dividing the stellar distribution along the chosen projection into 10 radial bins, such that there are an equal number of star particles in each bin. All galaxies have $>$ 80 star particles per bin, or $>$ 800 star particles within 1.5 $\times$ $R_{1/2}$, however the results do not vary significantly with bin choice. Interestingly, the value of the gradient does not appear to correlate with standard parameters such as halo mass, stellar mass, or \vmax (see appendix). However, there is a correlation with galaxy age. 

\begin{figure*}
	\includegraphics[width=0.45\textwidth, trim = 0 0 0 0]{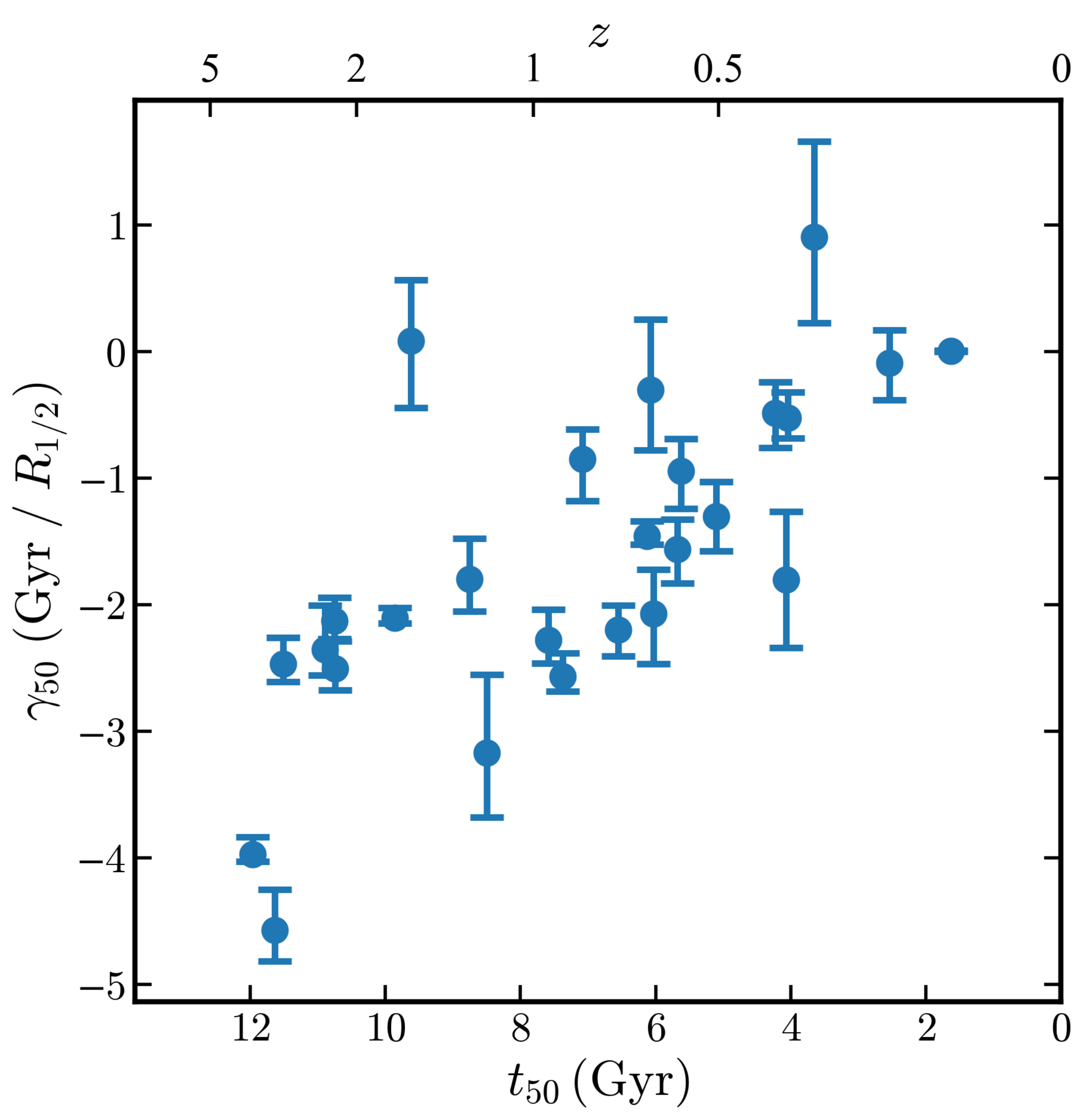}
	\includegraphics[width=0.45\textwidth, trim = 0 0 0 0]{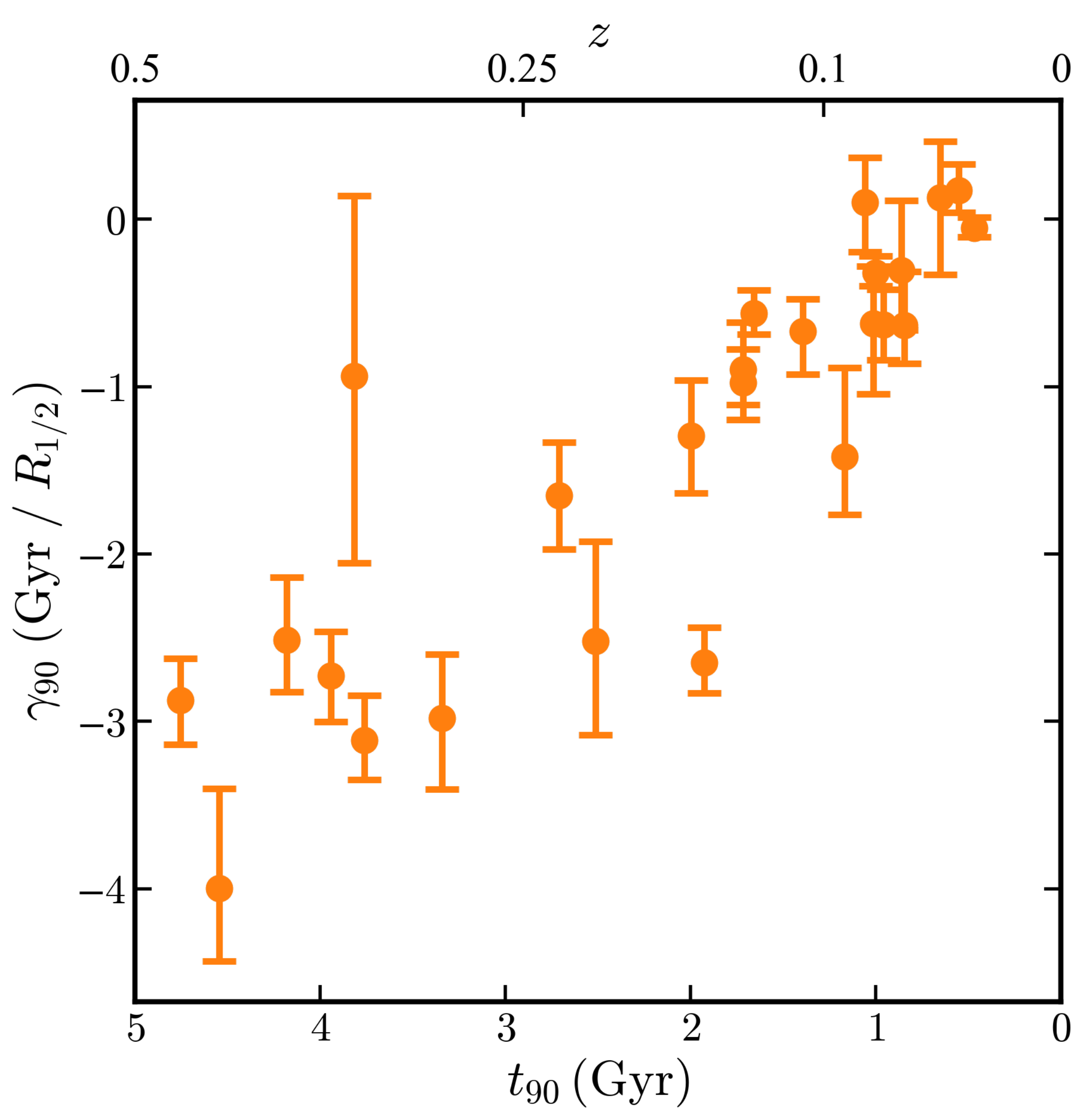}
	\centering
	\caption[Telescope-specific 3-color images]{Age gradients as defined in Equation \ref{eq:gamma_eq} for all of the galaxies in the sample, plotted against the $t_{50}$ (median age of stars, left) or $t_{90}$ (age of the youngest 10\% of stars, right) of all the stars in the galaxy.  Error bars represent the variation over all projections.   There is a clear trend with star formation time, such that galaxies that form earlier tend to have stronger gradients.}
	\label{fig:slope_time_relation}
\end{figure*}

Figure \ref{fig:slope_time_relation} shows the relationship between a galaxy's age gradient and the overall age of the galaxy. The error bars on the gradient values reflect the full variation over all projection angles while the ages are measured for all of the stars within 10\% of the virial radius. We see that earlier-forming galaxies show more significant (more negative) age gradients, while later-forming galaxies have smaller (less negative) age gradients.  We discuss the origin of this trend in Section \ref{sec:discussion} and conclude that it is driven by a combination of mergers and the strong stellar feedback inherent in our simulations.  

\section{Discussion}
\label{sec:discussion}

\begin{figure}
	\includegraphics[width=\linewidth, trim = 2.0cm 3.0cm 2.0cm 2.0cm]{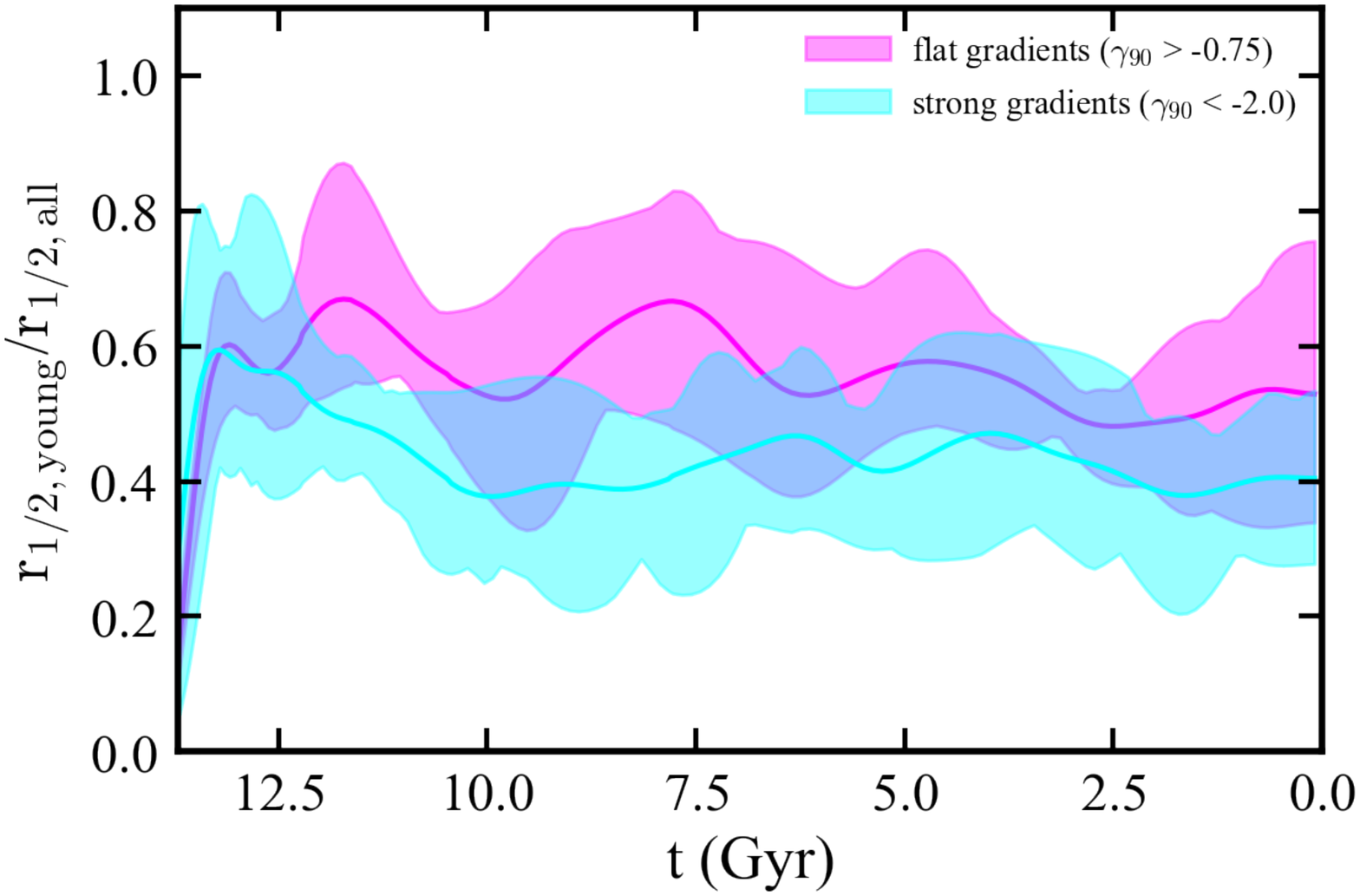}
	\centering
	\caption[radius_ratio]{The ratio between the 3D half-mass radius of young stars (r$_{\rm 1/2,young}$), and all of the stars in the galaxy at that same time (r$_{\rm 1/2,all}$) as a function of time. This sample is divided into two subsamples: galaxies with flat age gradients ($\gamma_{90}$ $\geq$ -0.75) and galaxies with strong age gradients ($\gamma_{90}$ $<$ -2.0). The two subsamples are then averaged and smoothed using a Gaussian kernel. Two things are apparent from this figure.  First, on average, stars are born near the center of the galaxy at all times, signaling that some process is heating the stars over time. Second, galaxies with flatter age gradients tend to form stars closer to the distribution of all stars, while galaxies with strong age gradients tend to have more concentrated star formation at nearly all times (with a potential exception at very early times).}
	\label{fig:radius_ratio}
\end{figure}

\subsection{Origin of Age Gradients}

The standard picture for how stars are distributed in galaxies as a function of age is rooted in the ``inside-out" model, where star formation is first confined to the center of the galaxy and proceeds in the outer reaches only at later times \citep{Larson76,Mo98,Avila00,Brook06,Pilkington12,Aumer13}. Such a distribution is seen observationally in larger galaxies including disk galaxies \citep{Martin05,Gil05,Gil07,Nelson12,Nelson16} and ellipticals \citep{vanDokkum10,Patel13}. Based on how age gradients are measured in this work, inside-out formation would correspond to a positive value of $\gamma_{50}$ and $\gamma_{90}$. From figure \ref{fig:slope_time_relation} we see that essentially none of the dwarf galaxies studied in this work show positive age gradients in a striking difference from that naive expectation.  

Unlike large disks and elliptical galaxies, dwarfs are much less ordered, and this likely contributes to the lack of the characteristic inside-out age gradients.  Two possible mechanisms for inverting age gradients are mergers, which could bring in stars and modify an existing gradient, and stellar feedback, which could heat stars progressively over time, pushing older stars to the outskirts.

Both galaxy mergers and mergers with dark matter halos containing no stars could drive SFH gradients. Specifically, mergers (either with stars or without) can heat the pre-existing stellar distribution \citep{Starkenburg15,Benitez16,Leaman17}. Since these mergers tend to happen at early times, they might be responsible for driving the spread with formation time as they would be more likely to heat older stars. 

Mergers with galaxies that contain stars can also drive age gradients by preferentially adding stars to the outskirts of a galaxy. This is similar to how the stellar halo of Milky Way-size galaxies is believed to be built \citep{Bullock05,Cooper10}. Since most mergers are with lower mass objects, most accretion events among dwarfs galaxies likely include many ultrafaint dwarf galaxies with very old stellar populations. Mergers of this kind could deposit old stars to the outskirts of the primary dwarfs.

In reality, galaxies experience both mergers and feedback. Therefore, the exact origin of a gradient in a galaxy must be evaluated on a case-by-case basis. However, some broad conclusions can be drawn from analysis of this suite of simulations. Merger history can play some part in the formation of an age gradient, however, it can both create and destroy an age gradient. For example, several galaxies with flatter age gradients (such as m10xe, m10xd, m10xd\_A, and m10xh) have late time mergers with another galaxy. If the merging galaxy has a sufficiently extended SFH, than the merger adds young stars to the outskirts and prevents an age gradient from forming. In contrast, some galaxies like m10xi have strong age gradients that form due to mergers. In the specific case of m10xi it has a merger with several small objects at z $\simeq$ 1 which bring in almost exclusively ancient stars and rapidly build an age gradient. The relationship between strength of the gradient and the specifics of mergers naturally explains the trend between strength of the age gradient and formation time. Objects that have mergers that bring in young stars at late times would have both late formation times and flat age gradients. While objects that have earlier mergers that bring in ancient stars would have earlier formation times and strong age gradients.

\begin{figure*}
	\includegraphics[width=0.95\textwidth, trim = 0 4cm 0 4cm]{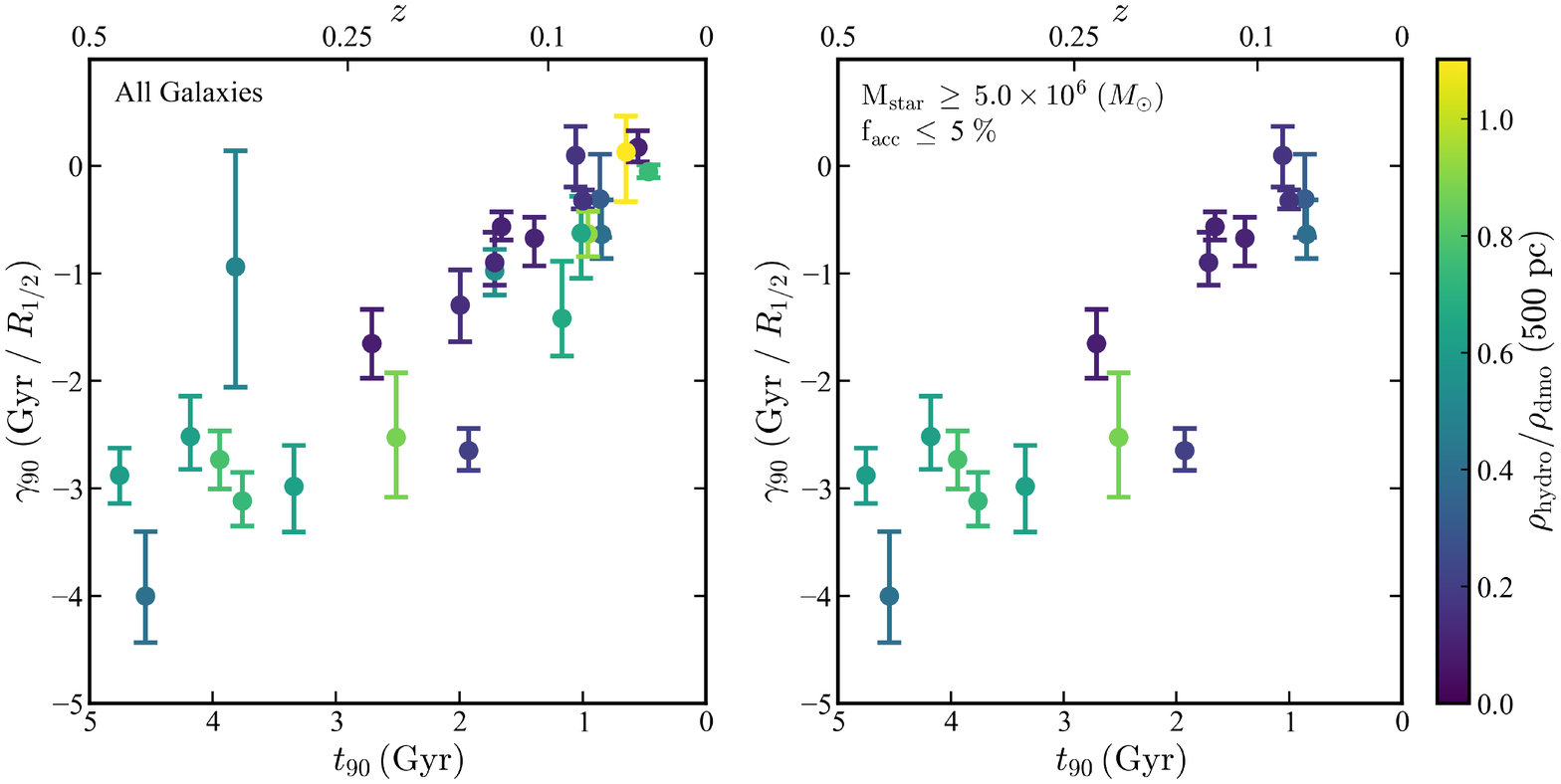}
    \includegraphics[width=0.95\textwidth, trim = 0 4cm 0 4cm]{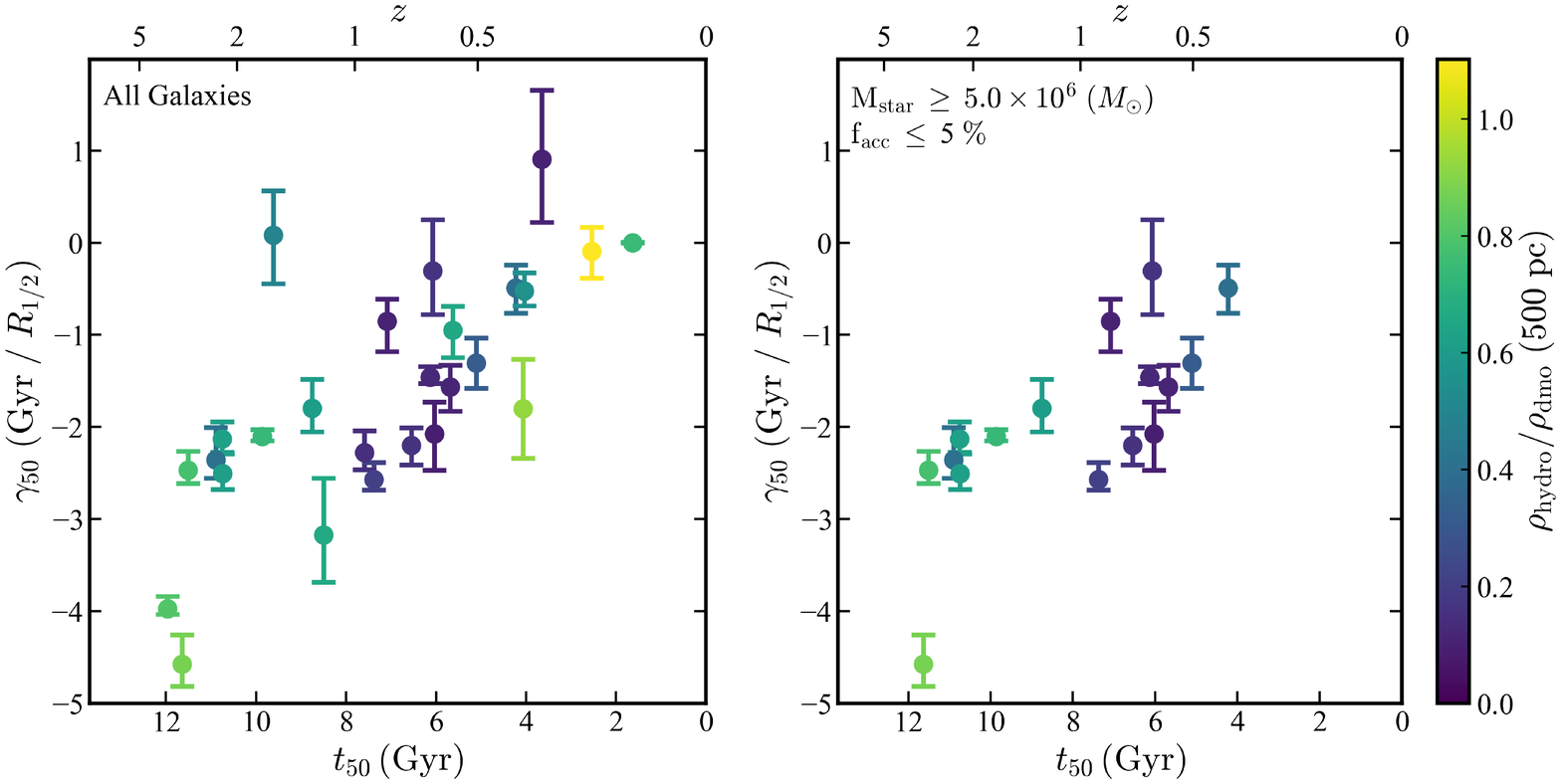}
	\centering
	\caption[core formation]{Strength of the age gradient as a function of formation time. The left two panels are identical to Figure \ref{fig:slope_time_relation}, however the points are now colored by the density ratio between the full hydrodynamics run and the corresponding halo from a dark matter only simulation. For a scenario where feedback is the only cause of the age gradients, one would expect a correlation between core formation and strength of gradient. However, galaxies with too low of a stellar mass to form a core, and galaxies with many stars brought in by mergers could contaminate this relationship. The right panels attempt to remove these contaminants, by removing any galaxy with less than 5$\times10^6$ $M_{\odot}$ in stellar mass, and with greater than 5\% accreted stars. Once these are removed there appears to be a slight relation between core formation and strength of the gradient, where objects flat gradients tend to have recent star formation, and large cores. This potentially points toward these stellar populations being more well mixed due to recent star bursts driving feedback and both creating cores, and smoothing out an age gradient.}
	\label{fig:core_formation}
\end{figure*}

Interestingly, most of the galaxies in our sample that have no accreted stars still have pronounced age gradients.  Evidence of this can be seen in Figure \ref{fig:radius_ratio} where we show the ratio between the half-mass radii of young stars (born since the last snapshot) and the half-mass radii of all stars. At all times, the half-mass radii of young stars is on average about half that of all stars, pointing to some mechanism that heats the stars after they form, pushing them to larger radii. Such a mechanism would build a gradient over time as successive generations of stars are pushed further out.

One obvious physical mechanism for creating this heating is stellar feedback processes similar to those that are responsible for creating cores in dwarf galaxies \citep{Pontzen12,DiCintio14,Onorbe15,Dutton16,Read16,Fitts17}. Stellar feedback from supernovae can quickly expel gas from a dwarf galaxy.  The resultant change in the gravitational potential can perturb the orbits of dark matter particles and create cored dark matter halo profiles. A similar process could reorder the stars in the galaxy.  One obvious indication that this is happening would be if the strength of the age gradient correlates in some way with the formation of a core in the dark matter halo.

As seen in previous works involving dwarf galaxies in FIRE \citep{Onorbe15,ElBadry16,Fitts17} the dark matter halos of dwarf galaxies in our sample can be strongly affected by stellar feedback, with cores getting larger as the stellar mass increases, up to the limit of our sample. Naturally, if stellar feedback were responsible for creating age gradients as well as cores, a first-order test would be to see if the strength of the gradient correlates with the size of the core. To measure core strength we ran dark matter only versions of our simulations and have compared the densities near the center both with and without hydrodynamics. In figure \ref{fig:core_formation} we plot the same data as figure \ref{fig:slope_time_relation}, but now color the points by the ratio of the dark matter density in the full hydrodynamics simulation to that of the corresponding dark matter only simulation ($\rho_{\rm hydro}$/$\rho_{\rm dmo}$) measured at 500 pc. The left panels in figure \ref{fig:core_formation} show this for all galaxies in our sample and there appears to be no relation between core formation and age gradient. However, this sample includes galaxies with late time mergers, which could flatten out the age gradient, but may have a different impact on the core. Furthermore, some of the smaller galaxies in our sample do not have enough stars for feedback to have affected the core \citep[e.g.][]{Fitts17}. Both of these mechanisms would act to wash out any obvious relation between core formation and gradient formation. 

In the right-hand panel of figure \ref{fig:core_formation} we plot the same relation, but only for galaxies that have enough stellar mass to form a core ($M_{\rm star}$ $>$ $5 \times 10^6$ M$_{\odot}$), and have few accreted stars ($f_{\rm acc}$ $\leq$ 5 \%).  For these galaxies we see some indication that galaxies with larger cores have {\it flatter} age gradients. This trend can be understood in tandem with the relationship between gradient strength and formation time. Early-forming galaxies have more feedback at early times to push out stars.  Subsequent star formation deposits young stars at small radius, creating a strong age gradient. However, any core in the dark matter halo that formed along with this early star formation can get rebuilt from subsequent dark-matter mergers, an effect that has been seen in previous FIRE simulations \citep{Onorbe15}. Subsequent star formation was not enough to drive the creation of a core, nor to drive younger stars out, but enough to drive the median age of the inner regions to younger ages. For later forming galaxies, large star formation rates at late times created a large core, and pushed younger stars into the outskirts mixing the stars of different ages at larger radii. Completely decoupling all these effects would require a much larger sample of galaxies with a variety of accretion histories at fixed stellar mass. Investigations of this nature with a larger sample of dwarf galaxies could help to decouple the timescales of core formation, gradient formation, and mergers from galaxies.

As a final caveat, several recent works have investigated the connection between core formation in dwarf galaxies and the density threshold of star formation implemented in cosmological simulations of galaxy formation \citep{Bose18,Dutton18,Benitez18}. In general, it appears as though core formation is intimately tied to the density threshold of star formation assumed. If the density threshold is too low ($\it{n}$ $\simeq 0.1 \,{\rm cm}^{-3}$), gas is allowed to turn into stars a low densities compared to the local dark matter density, and cannot dominate the local gravitational potential, preventing the formation of cores. On the other hand, if the density threshold is high ($\it{n}$ $\simeq$ 10-100 $cm^{-3}$), the gas can dominate the local gravitational potential. The density threshold for star formation used in our FIRE-2 simulations is ${\it n}=1000\,{\rm cm}^{-3}$, which is well into the regime where star formation is able to induce core formation. The relation between star formation threshold and core formation complicates the picture of how star formation impacts galaxy evolution. If the formation of age gradients is tied to feedback, then it could also be similarly affected by the threshold of star formation. If so, then observational explorations of age gradients in dwarf galaxies may provide an interesting direct constraint on galaxy formation simulations and help define realistic star formation thresholds.

\subsection{Observed Dwarf Galaxies}

It has long been known that the classical dwarf satellites of the Milky Way and galaxies throughout the Local Volume show metallicity gradients \citep{Harbeck01,Battaglia06,Bernard08,McConnachie12,Martinez15,Okamoto17,Kacharov17}. The metallicity gradients appear to be in the same direction of those in simulated galaxies presented in this work. Observed dwarf galaxies tend to have  metal-rich (younger) stars in their central regions, and metal-poor (older) stars farther out. To date, observations of the distribution of stellar ages as a function of radius, are rare, particularly at this mass scale, making a direct comparison difficult.

One comparison we can make is to the data presented in \cite{Hidalgo13}, who observed four nearby dwarf galaxies (Cetus, Tucana, LGS 3, and Phoenix) and measured the SFHs of these galaxies in radial bins. The results once again qualitatively agree with our results. LGS 3 and Phoenix show a fairly strong spread in SFH where the age of the latest forming stars increases with increasing radius. Tucana shows a smaller spread in its SFH, and Cetus has no detectable spread. It is difficult to use this data to quantify whether or not the formation time is correlated to the spread in the SFH because many of these galaxies are not isolated, and are therefore may be quenched by environmental effects \citep{Fillingham18}. Indeed, the galaxies with smallest gradients in the \citet{Hidalgo13} sample  (Cetus and Tucana) are older; however, both of these galaxies are quenched whereas all the isolated dwarf galaxies in FIRE studied in this paper are actively star forming. Another comparison that can be made is to various Integral Field Unit surveys of galaxies that have become common over the past several years. Because these surveys provide spectral coverage over each pixel, it is possible to measure SFH variation over the face of a galaxy. 

\citet{GarciaBenito17} measured the spatially resolved SFHs for 661 galaxies within the CALIFA survey. The stellar mass range of their targeted galaxies is $10^{8.4}$ to $10^{12}$ M$_{\odot}$, and thus only the most massive of the galaxies in our sample are directly comparable to the results from the CALIFA survey. However, there are some interesting trends that are hinted at in \citet{GarciaBenito17}. In general, they find that their galaxies show age gradients with older stars in the central regions, and younger stars at the outskirts consistent with typical inside-out galaxy formation. These massive galaxies also show very weak gradients in their SFHs. Interestingly, this gradient actually strengthens in the lowest-mass bin in their sample, $M_{\rm star}$ = $10^{8.4}$ to $10^{9.9}$ M$_{\odot}$. In this bin, the stars are in general later forming, but the gradient between the inner and outermost regions is larger. This could potentially point toward some interesting evolution between the most massive galaxies in our sample which show slightly negative or flat gradients, and the lowest mass galaxies in the CALIFA sample, which show large positive age gradients. Additionally, it is possible that there is a strong difference between SFHs measured from spectra, and those measured from CMDs. Such a tension was pointed out by \citet{Leitner12} who compared SFH measured with both SDSS spectra and CMDs and found a that they disagree, with SFHs from CMDs implying a much slower growth. However, in order to decouple these issues, more observations of spatially resolved SFHs for small galaxies are needed.

\begin{figure*}
	\includegraphics[width=0.45\textwidth, trim = 0 0 0 0]{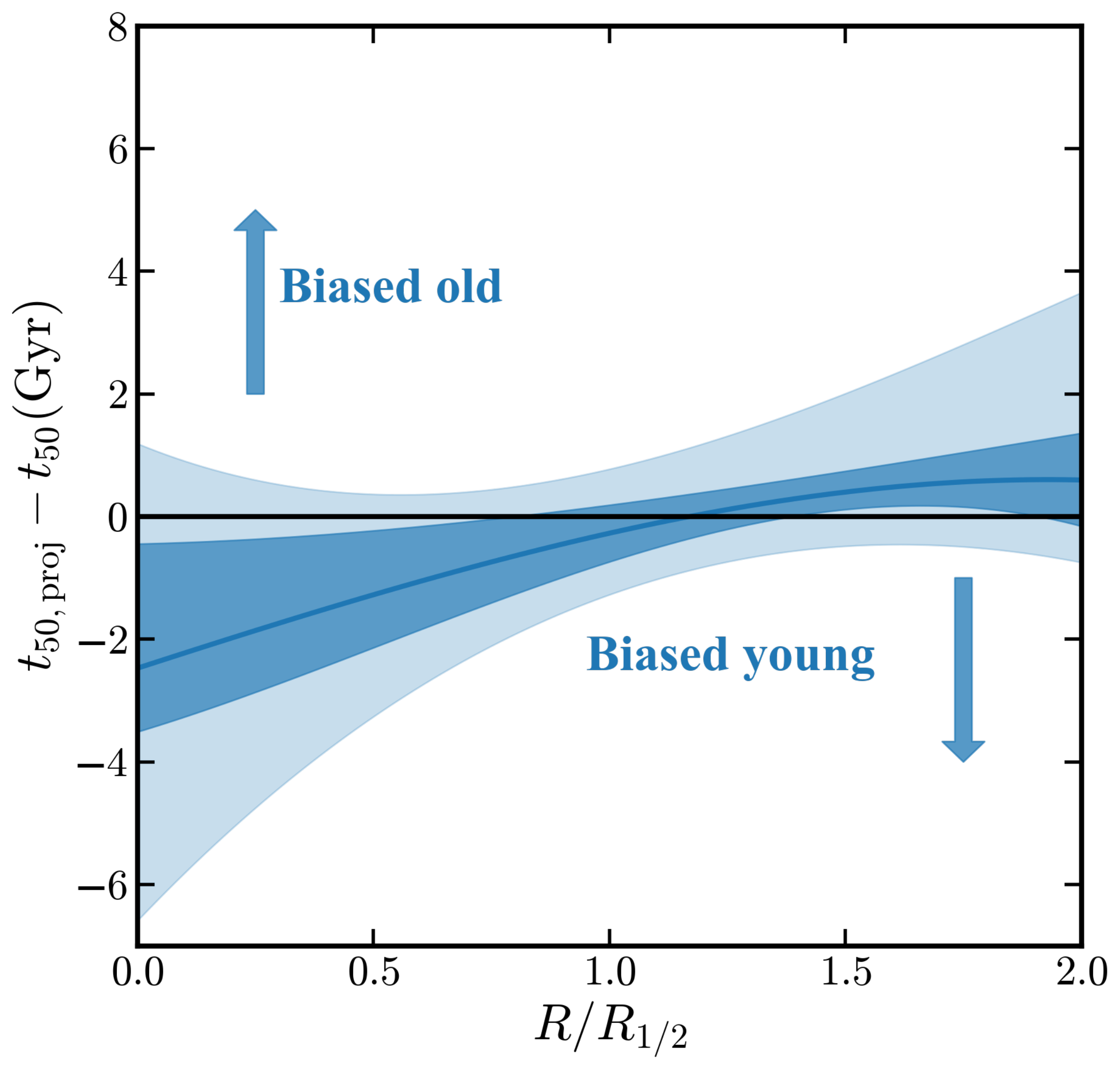}
	\includegraphics[width=0.45\textwidth, trim = 0 0 0 0]{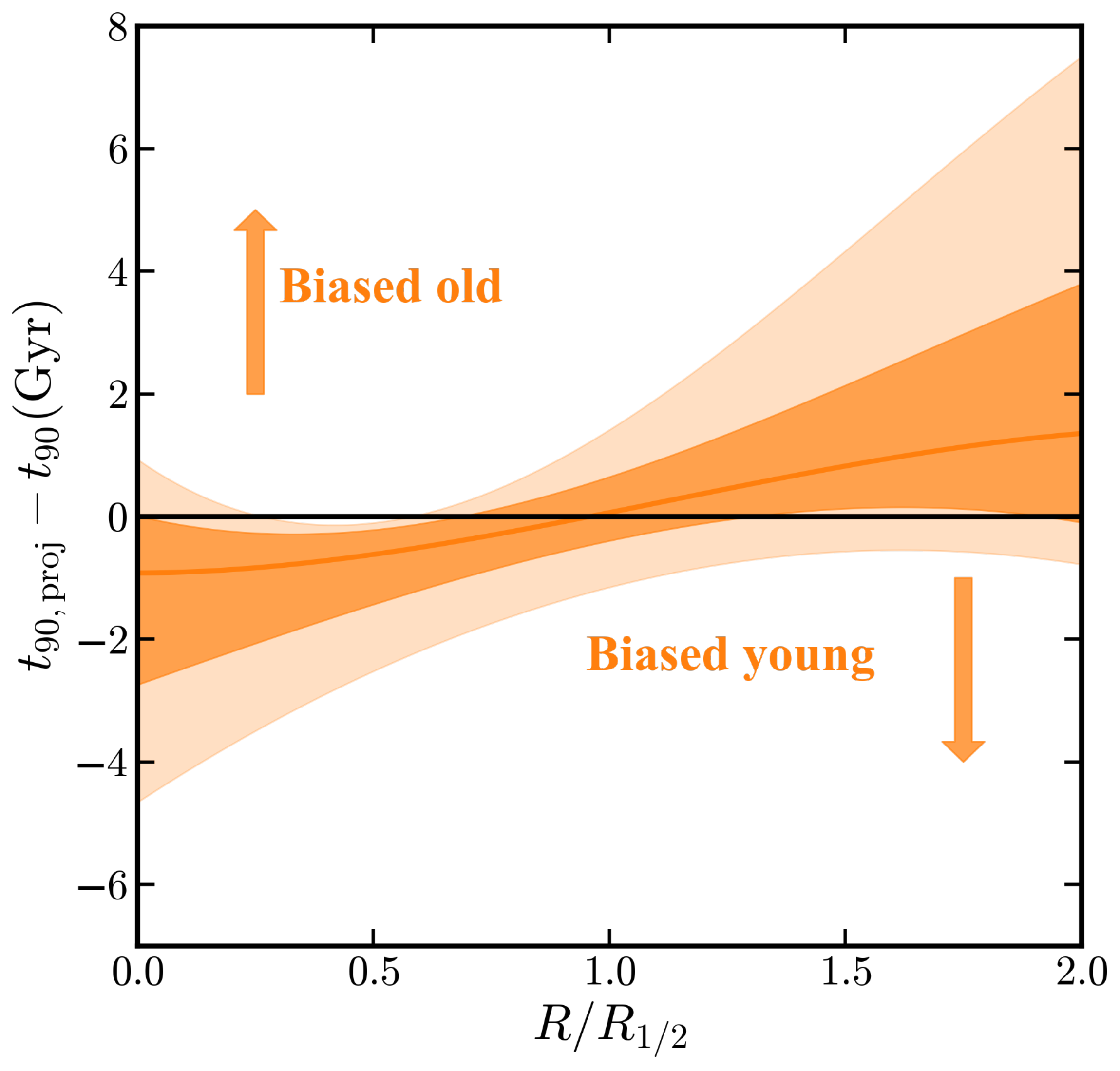}
	\centering
	\caption[Bias]{Bias in inferred SFH that would occur from studying a stellar population at a given projected radius in units of a galaxy's half-mass radius.  The offset in true \tfifty (median age) and \tninty (90 percentile youngest stars) are shown in the left and right panels, respectively. The solid lines represent the median bias in age determined by ``stacking" all of our galaxies and observing them over 100 viewing angles each. The darker and lighter shaded regions represent the 68\% and 95\% contours of the distributions.  The ``pinch" at a time-bias of zero at $R/R_{1/2} 
    \simeq 1$ in both panels shows that photometric fields positioned near a galaxy's half-light radius are optimal for single-pointing stellar population studies aimed at providing an unbiased view of a galaxy's global SFH.}
	\label{fig:error_with_r}
\end{figure*}

\subsection{Implications for SFH measurements}
\label{sec:bias}

If real dwarf galaxies have SFH gradients similar to those predicted in our simulations, it could impact the interpretation of current measurements of SFHs and plans for future measurements. This is because CMD-derived SFHs often rely on photometric observations via space-based facilities, such as the {\it Hubble Space Telescope} ({\it HST}), with fields of view that cover only 10\% (or less) of the area of nearby galaxies. If the galaxy has a strong age gradient, the placement of the field with respect to the galaxy's center could lead to a biased view of the overall global SFH of the galaxy.  

Here we explore the potential bias that could arise from fractional spatial coverage by measuring the spread in SFHs as a function of projected radius. Of course any bias in the recovered SFH depends on several factors including the survey field's distance from the center of the galaxy, the size of the field of view with respect to the galaxy, and the steepness of the gradient. Observationally, the first two are known, but the strength of the gradient cannot be determined without first measuring the SFH. Our goal is to estimate the nature and magnitude of the effect using mock observations of our galaxy sample.

Figure \ref{fig:error_with_r} shows the result of this exercise.  We have computed the values of \tfifty and \tninty for all galaxies in our sample, each viewed from 100 different random angles. We then measured the spread in SFH in discrete bins of  width $0.1 \, R_{1/2}$. The figure shows the measured ages relative to the true value of \tfifty or \tninty for all of the stars within 10\% of the virial radius of the galaxy's dark matter halo. The solid lines in Figure \ref{fig:error_with_r} show the median of the error in age as a function of projected radius.  The shaded regions with the darker and lighter bands show the 68\% and 95\% distributions in deviation from the true age of the galaxy.

The bias in the SFH measured at different points throughout the galaxy works in the direction one would expect: the inner regions of the galaxy are biased younger than the global SFH of the galaxy, and the outskirts are biased older than the global value. Interestingly, the SFH measured around the projected half mass radius matches well to the galaxy-averaged value, with a relatively small scatter. We can apply this information to current observed SFHs such as those in \cite{Weisz14}. In general, this data is archival {\it HST} data, where the fields were selected to maximize the number of stars. Therefore, the fields tend to be near the centers of galaxies, and are almost always within the half-light radius. Assuming the half-mass and half-light radii are comparable in size, this would imply that many CMD-based SFHs in \cite{Weisz14} are biased young relative to the global SFH of the galaxy as mentioned in section 4.1 of \cite{Weisz14}. This motivates additional observations of these galaxies outside of the half-light radii.

\subsection{Implications for future observations}

As discussed in Section \ref{sec:bias}, age gradients can potentially bias SFHs inferred from CMD studies that cover only a fraction of a dwarf galaxy. This result may have important implications for planning future observing campaigns to infer galaxy SFHs with resolved stellar populations. Without a careful accounting for the age gradient, significant errors in the SFH can result. Where fields should be placed within galaxies in order to recover the correct SFH will become even more important in the future for the {\it James Webb Space Telescope} ({\it JWST}), which has a similar field of view as {\it HST} but will be able to measure SFHs for many more galaxies throughout the Local Volume. Our results shown in Figure \ref{fig:error_with_r} suggest that fields at radii close to $R_{1/2}$ are optimal. However, for galaxies of these sizes beyond about 1.3 Mpc the entire galaxy should fall within one {\it HST} or {\it JWST} pointing, thus eliminating the need to aim for $R_{1/2}$. Note that we find no significant spatial or angular variation at fixed projected radius -- it is the distance from the galaxy center that matters most.  

Even further on the in the future, WFIRST will be able to cover most of the area of galaxies throughout the Local Volume; however, crowding effects could limit its ability to measure accurate SFHs for the inner regions of these galaxies.  If crowding affects the ability to do stellar population studies within $R_{1/2}$, then measured SFHs could be biased old with respect to the global stellar distribution. 

One potential caveat of the analysis in this section concerns whether the simulated galaxies we study here are representative of dwarf galaxies in the Local Group.  The relative isolation of the galaxies in our sample could make their SFHs differ from dwarf galaxies in the vicinity of a larger galaxy. One place where our sample is truly not representative is to the satellites of the MW and M31, where most low-mass galaxies are not actively star forming. However, it is possible that host interactions influence dwarf galaxy evolution even beyond the virial radii of M31 or the MW. Garrison-Kimmel et al. (in prep) have looked at this question by comparing dwarf galaxies simulated with FIRE physics.  They concluded that the isolated dwarfs used in this work form significantly later than satellite galaxies at all masses. Interestingly, they also form slightly later on average than galaxies that are beyond the virial radius of a larger system but within about 1 Mpc of a Local Group giant. This difference is not statistically significant for galaxies of a stellar mass above $10^6$ M$_{\odot}$, which all but two of the galaxies used in this sample are. 

\section{Conclusion} \label{s:conclusion}

In this work we have studied the presence of SFH gradients in FIRE-2 simulations of dwarf galaxies. Almost all of our galaxies, which have stellar masses between $10^{5.5}$ and $10^{8.6}$ M$_{\odot}$, show gradients in their SFHs with younger stars at the center and older stars at the outskirts. The slope of the gradients varies widely between galaxies, and does not seem to show a strong correlation with stellar mass and halo mass. However, as shown in Figure \ref{fig:slope_time_relation}, there is a clear correlation between the slope of the age gradient and the overall age of the galaxy.   

The origin of these age gradients is a complicated mixture of both mergers and stellar feedback. Mergers can both create and destroy age gradients based on the timing of a merger, and the SFH of the merging system. Stellar feedback produces potential fluctuations that heat stars and also launches molecular outflows that subsequently form stars \citep{ElBadry16}.  Earlier forming galaxies have had a longer time for their stars to be dynamically heated by multiple feedback episodes changing the gravitational potential of the galaxy via the removal of gas.  Early feedback events also produce more outflowing, star-forming gas, which eventually deposits older stars at larger radii.

Furthermore, we see a trend between strength of the age gradient and core formation in the dark matter halo. Systems with flat age gradients tend to have large dark matter cores. This is because these systems have had more recent intense star formation which both builds a core and efficiently mixes stellar populations of different ages. For galaxies with strong age gradients, these systems tend to have stronger star formation at earlier times, and over time these stars were heated by stellar feedback. At late times the SFR was low enough that feedback could not drive a core in the dark matter halo, or eject stars into the outskirts of the galaxy, but was enough to drive the median age of the central regions to younger ages. 

Observational comparisons to available data are difficult; however, there are some broad consistencies. First, dwarf galaxies are known to have median age gradients with younger stars at the center and older stars at the outskirts. Resolved SFH work such as that from \citet{Hidalgo13} supports the existence of SFH gradients, but provides weak evidence of a trend with formation time. 

The existence of significant age gradients in dwarf galaxies could potentially lead to observational biases in stellar population studies that are restricted to a limited area of a dwarf.  Figure \ref{fig:error_with_r} shows that inner fields are biased young and outer fields are biased old relative to the global SFH of the galaxy.  The ideal location for a photometric field to provide an unbiased measure of the global SFH is near the 2D half-light radius. 

The observation of SFH gradients in dwarf galaxies has several implications for the future of galaxy formation studies. If dwarf galaxies in the Local Volume show a gradient in their SFHs, measuring how this gradient scales with formation time could be useful in calibrating feedback models. However, accurate calibration of feedback models would require resolved SFHs for many dwarf galaxies at several radii in order to accurately measure the slopes of the gradient.

Importantly, the spread in SFH with radius is large enough that SFHs derived from single-field CMDs may not be accurate. The SFH can vary significantly with radius, potentially biasing results from small fields by several Gyrs. Thus truly understanding the formation of dwarf galaxies will require observations covering large areas of the galaxy, or at least some correction for how incomplete coverage can bias measurements of the SFH. If only a single pointing is possible then Figure \ref{fig:error_with_r} suggests that the field should be centered near $R_{1/2}$. Such considerations should be taken into account as we move towards stellar population studies in the era of {\it JWST}, {\it WFIRST} and 30m-class telescopes.

\section{Acknowledgments} \label{s:Acknowledgements}

ASG and JSB were supported by NSF AST-1518291, HST-AR-14282, and HST-AR-13888. ASG is further supported by the McDonald Observatory at the University of Texas at Austin, through the Harlan J. Smith fellowship. MBK and AF acknowledge support from NSF grant AST-1517226. MBK also acknowledges support from NSF CAREER grant AST-1752913 and from NASA grants NNX17AG29G and HST-AR-13888, HST-AR-13896, HST-AR-14282, HST-AR-14554, HST-AR-15006, HST-GO-12914, and HST-GO-14191 from the Space Telescope Science Institute, which is operated by AURA, Inc., under NASA contract NAS5-26555. MCC was supported by NSF AST-1815475. DRW is supported by a fellowship from the Alfred P. Sloan Foundation. He also acknowledges support from the Alexander von Humboldt Foundation and from HST-GO-13768, HST-GO-15476, HST-AR-13888, HST-AR-13925, HST-AR-15006, and JWST-ERS-01334. AW was supported by NASA through ATP grant 80NSSC18K1097 and grants HST-GO-14734 and HST-AR-15057 from STScI. RF acknowledges support from the Swiss National Science Foundation (grant no. 157591). CAFG was supported by NSF through grants AST-1517491, AST-1715216, and CAREER award AST-1652522, by NASA through grants NNX15AB22G and 17-ATP17-0067, and by a Cottrell Scholar Award from the Research Corporation for Science Advancement. This work used computational resources of the University of Texas at Austin and the Texas Advanced Computing Center (TACC; http://www.tacc.utexas.edu), the NASA Advanced Supercomputing (NAS) Division and the NASA Center for Climate Simulation (NCCS), and the Extreme Science and Engineering Discovery Environment (XSEDE), which is supported by National Science Foundation grant number OCI-1053575.



\bibliographystyle{mnras}
\bibliography{SFH_paper_refs.bbl} 

\begin{thebibliography}{}
\makeatletter
\relax
\def\mn@urlcharsother{\let\do\@makeother \do\$\do\&\do\#\do\^\do\_\do\%\do\~}
\def\mn@doi{\begingroup\mn@urlcharsother \@ifnextchar [ {\mn@doi@}
  {\mn@doi@[]}}
\def\mn@doi@[#1]#2{\def\@tempa{#1}\ifx\@tempa\@empty \href
  {http://dx.doi.org/#2} {doi:#2}\else \href {http://dx.doi.org/#2} {#1}\fi
  \endgroup}
\def\mn@eprint#1#2{\mn@eprint@#1:#2::\@nil}
\def\mn@eprint@arXiv#1{\href {http://arxiv.org/abs/#1} {{\tt arXiv:#1}}}
\def\mn@eprint@dblp#1{\href {http://dblp.uni-trier.de/rec/bibtex/#1.xml}
  {dblp:#1}}
\def\mn@eprint@#1:#2:#3:#4\@nil{\def\@tempa {#1}\def\@tempb {#2}\def\@tempc
  {#3}\ifx \@tempc \@empty \let \@tempc \@tempb \let \@tempb \@tempa \fi \ifx
  \@tempb \@empty \def\@tempb {arXiv}\fi \@ifundefined
  {mn@eprint@\@tempb}{\@tempb:\@tempc}{\expandafter \expandafter \csname
  mn@eprint@\@tempb\endcsname \expandafter{\@tempc}}}

\bibitem[\protect\citeauthoryear{{Aumer} \& {White}}{{Aumer} \&
  {White}}{2013}]{Aumer13}
{Aumer} M.,  {White} S.~D.~M.,  2013, \mn@doi [\mnras] {10.1093/mnras/sts083},
  \href {http://adsabs.harvard.edu/abs/2013MNRAS.428.1055A} {428, 1055}

\bibitem[\protect\citeauthoryear{{Avila-Reese} \& {Firmani}}{{Avila-Reese} \&
  {Firmani}}{2000}]{Avila00}
{Avila-Reese} V.,  {Firmani} C.,  2000, \rmxaa, \href
  {http://adsabs.harvard.edu/abs/2000RMxAA..36...23A} {36, 23}

\bibitem[\protect\citeauthoryear{{Battaglia} et~al.,}{{Battaglia}
  et~al.}{2006}]{Battaglia06}
{Battaglia} G.,  et~al., 2006, \mn@doi [\aap] {10.1051/0004-6361:20065720},
  \href {http://adsabs.harvard.edu/abs/2006A%26A...459..423B} {459, 423}

\bibitem[\protect\citeauthoryear{{Beccari} et~al.,}{{Beccari}
  et~al.}{2014}]{Beccari14}
{Beccari} G.,  et~al., 2014, \mn@doi [\aap] {10.1051/0004-6361/201424411},
  \href {http://adsabs.harvard.edu/abs/2014A%26A...570A..78B} {570, A78}

\bibitem[\protect\citeauthoryear{{Behroozi}, {Wechsler}  \& {Wu}}{{Behroozi}
  et~al.}{2013}]{Behroozi13}
{Behroozi} P.~S.,  {Wechsler} R.~H.,   {Wu} H.-Y.,  2013, \mn@doi [\apj]
  {10.1088/0004-637X/762/2/109}, \href
  {http://adsabs.harvard.edu/abs/2013ApJ...762..109B} {762, 109}

\bibitem[\protect\citeauthoryear{{Benitez-Llambay}, {Navarro}, {Abadi},
  {Gottl{\"o}ber}, {Yepes}, {Hoffman}  \& {Steinmetz}}{{Benitez-Llambay}
  et~al.}{2016}]{Benitez16}
{Benitez-Llambay} A.,  {Navarro} J.~F.,  {Abadi} M.~G.,  {Gottl{\"o}ber} S.,
  {Yepes} G.,  {Hoffman} Y.,   {Steinmetz} M.,  2016, \mn@doi [\mnras]
  {10.1093/mnras/stv2722}, \href
  {http://adsabs.harvard.edu/abs/2016MNRAS.456.1185B} {456, 1185}

\bibitem[\protect\citeauthoryear{{Benitez-Llambay}, {Frenk}, {Ludlow}  \&
  {Navarro}}{{Benitez-Llambay} et~al.}{2018}]{Benitez18}
{Benitez-Llambay} A.,  {Frenk} C.~S.,  {Ludlow} A.~D.,   {Navarro} J.~F.,
  2018, preprint, \href {http://adsabs.harvard.edu/abs/2018arXiv181004186B} {}
  (\mn@eprint {arXiv} {1810.04186})

\bibitem[\protect\citeauthoryear{{Bernard} et~al.,}{{Bernard}
  et~al.}{2008}]{Bernard08}
{Bernard} E.~J.,  et~al., 2008, \mn@doi [\apjl] {10.1086/588285}, \href
  {http://adsabs.harvard.edu/abs/2008ApJ...678L..21B} {678, L21}

\bibitem[\protect\citeauthoryear{{Bose} et~al.,}{{Bose} et~al.}{2018}]{Bose18}
{Bose} S.,  et~al., 2018, preprint, \href
  {http://adsabs.harvard.edu/abs/2018arXiv181003635B} {} (\mn@eprint {arXiv}
  {1810.03635})

\bibitem[\protect\citeauthoryear{{Boylan-Kolchin}, {Bullock}  \&
  {Kaplinghat}}{{Boylan-Kolchin} et~al.}{2011}]{MBK2011}
{Boylan-Kolchin} M.,  {Bullock} J.~S.,   {Kaplinghat} M.,  2011, \mn@doi
  [\mnras] {10.1111/j.1745-3933.2011.01074.x}, \href
  {http://adsabs.harvard.edu/abs/2011MNRAS.415L..40B} {415, L40}

\bibitem[\protect\citeauthoryear{{Boylan-Kolchin}, {Bullock}  \&
  {Garrison-Kimmel}}{{Boylan-Kolchin} et~al.}{2014}]{MBK2014}
{Boylan-Kolchin} M.,  {Bullock} J.~S.,   {Garrison-Kimmel} S.,  2014, \mn@doi
  [\mnras] {10.1093/mnrasl/slu079}, \href
  {http://adsabs.harvard.edu/abs/2014MNRAS.443L..44B} {443, L44}

\bibitem[\protect\citeauthoryear{{Boylan-Kolchin}, {Weisz}, {Johnson},
  {Bullock}, {Conroy}  \& {Fitts}}{{Boylan-Kolchin} et~al.}{2015}]{MBK2015}
{Boylan-Kolchin} M.,  {Weisz} D.~R.,  {Johnson} B.~D.,  {Bullock} J.~S.,
  {Conroy} C.,   {Fitts} A.,  2015, \mn@doi [\mnras] {10.1093/mnras/stv1736},
  \href {http://adsabs.harvard.edu/abs/2015MNRAS.453.1503B} {453, 1503}

\bibitem[\protect\citeauthoryear{{Brook}, {Kawata}, {Martel}, {Gibson}  \&
  {Bailin}}{{Brook} et~al.}{2006}]{Brook06}
{Brook} C.~B.,  {Kawata} D.,  {Martel} H.,  {Gibson} B.~K.,   {Bailin} J.,
  2006, \mn@doi [\apj] {10.1086/499154}, \href
  {http://adsabs.harvard.edu/abs/2006ApJ...639..126B} {639, 126}

\bibitem[\protect\citeauthoryear{{Brown} et~al.,}{{Brown}
  et~al.}{2014}]{Brown14}
{Brown} T.~M.,  et~al., 2014, \mn@doi [\apj] {10.1088/0004-637X/796/2/91},
  \href {http://adsabs.harvard.edu/abs/2014ApJ...796...91B} {796, 91}

\bibitem[\protect\citeauthoryear{{Bullock} \& {Boylan-Kolchin}}{{Bullock} \&
  {Boylan-Kolchin}}{2017}]{BBK17}
{Bullock} J.~S.,  {Boylan-Kolchin} M.,  2017, \mn@doi [\araa]
  {10.1146/annurev-astro-091916-055313}, \href
  {http://adsabs.harvard.edu/abs/2017ARA%26A..55..343B} {55, 343}

\bibitem[\protect\citeauthoryear{{Bullock} \& {Johnston}}{{Bullock} \&
  {Johnston}}{2005}]{Bullock05}
{Bullock} J.~S.,  {Johnston} K.~V.,  2005, \mn@doi [\apj] {10.1086/497422},
  \href {http://adsabs.harvard.edu/abs/2005ApJ...635..931B} {635, 931}

\bibitem[\protect\citeauthoryear{{Bullock}, {Kravtsov}  \&
  {Weinberg}}{{Bullock} et~al.}{2000}]{Bullock00}
{Bullock} J.~S.,  {Kravtsov} A.~V.,   {Weinberg} D.~H.,  2000, \mn@doi [\apj]
  {10.1086/309279}, \href {http://adsabs.harvard.edu/abs/2000ApJ...539..517B}
  {539, 517}

\bibitem[\protect\citeauthoryear{{Cicu{\'e}ndez} \&
  {Battaglia}}{{Cicu{\'e}ndez} \& {Battaglia}}{2018}]{Cicuendez18b}
{Cicu{\'e}ndez} L.,  {Battaglia} G.,  2018, \mn@doi [\mnras]
  {10.1093/mnras/sty1748}, \href
  {http://adsabs.harvard.edu/abs/2018MNRAS.480..251C} {480, 251}

\bibitem[\protect\citeauthoryear{{Cicu{\'e}ndez} et~al.,}{{Cicu{\'e}ndez}
  et~al.}{2018}]{Cicuendez18a}
{Cicu{\'e}ndez} L.,  et~al., 2018, \mn@doi [\aap]
  {10.1051/0004-6361/201731450}, \href
  {http://adsabs.harvard.edu/abs/2018A%26A...609A..53C} {609, A53}

\bibitem[\protect\citeauthoryear{{Cole}, {Weisz}, {Dolphin}, {Skillman},
  {McConnachie}, {Brooks}  \& {Leaman}}{{Cole} et~al.}{2014}]{Cole14}
{Cole} A.~A.,  {Weisz} D.~R.,  {Dolphin} A.~E.,  {Skillman} E.~D.,
  {McConnachie} A.~W.,  {Brooks} A.~M.,   {Leaman} R.,  2014, \mn@doi [\apj]
  {10.1088/0004-637X/795/1/54}, \href
  {http://adsabs.harvard.edu/abs/2014ApJ...795...54C} {795, 54}

\bibitem[\protect\citeauthoryear{{Cooper} et~al.,}{{Cooper}
  et~al.}{2010}]{Cooper10}
{Cooper} A.~P.,  et~al., 2010, \mn@doi [\mnras]
  {10.1111/j.1365-2966.2010.16740.x}, \href
  {http://adsabs.harvard.edu/abs/2010MNRAS.406..744C} {406, 744}

\bibitem[\protect\citeauthoryear{{Di Cintio}, {Brook}, {Macci{\`o}}, {Stinson},
  {Knebe}, {Dutton}  \& {Wadsley}}{{Di Cintio} et~al.}{2014}]{DiCintio14}
{Di Cintio} A.,  {Brook} C.~B.,  {Macci{\`o}} A.~V.,  {Stinson} G.~S.,  {Knebe}
  A.,  {Dutton} A.~A.,   {Wadsley} J.,  2014, \mn@doi [\mnras]
  {10.1093/mnras/stt1891}, \href
  {http://adsabs.harvard.edu/abs/2014MNRAS.437..415D} {437, 415}

\bibitem[\protect\citeauthoryear{{Dolphin}}{{Dolphin}}{2002}]{Dolphin02}
{Dolphin} A.~E.,  2002, \mn@doi [\mnras] {10.1046/j.1365-8711.2002.05271.x},
  \href {http://adsabs.harvard.edu/abs/2002MNRAS.332...91D} {332, 91}

\bibitem[\protect\citeauthoryear{{Dutton}, {Macci{\`o}}, {Frings}, {Wang},
  {Stinson}, {Penzo}  \& {Kang}}{{Dutton} et~al.}{2016}]{Dutton16}
{Dutton} A.~A.,  {Macci{\`o}} A.~V.,  {Frings} J.,  {Wang} L.,  {Stinson}
  G.~S.,  {Penzo} C.,   {Kang} X.,  2016, \mn@doi [\mnras]
  {10.1093/mnrasl/slv193}, \href
  {http://adsabs.harvard.edu/abs/2016MNRAS.457L..74D} {457, L74}

\bibitem[\protect\citeauthoryear{{Dutton}, {Macci{\`o}}, {Buck}, {Dixon},
  {Blank}  \& {Obreja}}{{Dutton} et~al.}{2018}]{Dutton18}
{Dutton} A.~A.,  {Macci{\`o}} A.~V.,  {Buck} T.,  {Dixon} K.~L.,  {Blank} M.,
  {Obreja} A.,  2018, preprint, \href
  {http://adsabs.harvard.edu/abs/2018arXiv181110625D} {} (\mn@eprint {arXiv}
  {1811.10625})

\bibitem[\protect\citeauthoryear{{Efstathiou}}{{Efstathiou}}{1992}]{Efstathiou92}
{Efstathiou} G.,  1992, \mn@doi [\mnras] {10.1093/mnras/256.1.43P}, \href
  {http://adsabs.harvard.edu/abs/1992MNRAS.256P..43E} {256, 43P}

\bibitem[\protect\citeauthoryear{{El-Badry}, {Wetzel}, {Geha}, {Hopkins},
  {Kere{\v s}}, {Chan}  \& {Faucher-Gigu{\`e}re}}{{El-Badry}
  et~al.}{2016}]{ElBadry16}
{El-Badry} K.,  {Wetzel} A.,  {Geha} M.,  {Hopkins} P.~F.,  {Kere{\v s}} D.,
  {Chan} T.~K.,   {Faucher-Gigu{\`e}re} C.-A.,  2016, \mn@doi [\apj]
  {10.3847/0004-637X/820/2/131}, \href
  {http://adsabs.harvard.edu/abs/2016ApJ...820..131E} {820, 131}

\bibitem[\protect\citeauthoryear{{Escala} et~al.,}{{Escala}
  et~al.}{2018}]{Escala18}
{Escala} I.,  et~al., 2018, \mn@doi [\mnras] {10.1093/mnras/stx2858}, \href
  {http://adsabs.harvard.edu/abs/2018MNRAS.474.2194E} {474, 2194}

\bibitem[\protect\citeauthoryear{{Faria}, {Feltzing}, {Lundstr{\"o}m},
  {Gilmore}, {Wahlgren}, {Ardeberg}  \& {Linde}}{{Faria}
  et~al.}{2007}]{Faria07}
{Faria} D.,  {Feltzing} S.,  {Lundstr{\"o}m} I.,  {Gilmore} G.,  {Wahlgren}
  G.~M.,  {Ardeberg} A.,   {Linde} P.,  2007, \mn@doi [\aap]
  {10.1051/0004-6361:20065244}, \href
  {http://adsabs.harvard.edu/abs/2007A%26A...465..357F} {465, 357}

\bibitem[\protect\citeauthoryear{{Faucher-Gigu{\`e}re}, {Lidz}, {Zaldarriaga}
  \& {Hernquist}}{{Faucher-Gigu{\`e}re} et~al.}{2009}]{Faucher09}
{Faucher-Gigu{\`e}re} C.-A.,  {Lidz} A.,  {Zaldarriaga} M.,   {Hernquist} L.,
  2009, \mn@doi [\apj] {10.1088/0004-637X/703/2/1416}, \href
  {http://adsabs.harvard.edu/abs/2009ApJ...703.1416F} {703, 1416}

\bibitem[\protect\citeauthoryear{{Fillingham}, {Cooper}, {Wheeler},
  {Garrison-Kimmel}, {Boylan-Kolchin}  \& {Bullock}}{{Fillingham}
  et~al.}{2015}]{Fillingham15}
{Fillingham} S.~P.,  {Cooper} M.~C.,  {Wheeler} C.,  {Garrison-Kimmel} S.,
  {Boylan-Kolchin} M.,   {Bullock} J.~S.,  2015, \mn@doi [\mnras]
  {10.1093/mnras/stv2058}, \href
  {http://adsabs.harvard.edu/abs/2015MNRAS.454.2039F} {454, 2039}

\bibitem[\protect\citeauthoryear{{Fillingham}, {Cooper}, {Pace},
  {Boylan-Kolchin}, {Bullock}, {Garrison-Kimmel}  \& {Wheeler}}{{Fillingham}
  et~al.}{2016}]{Fillingham16}
{Fillingham} S.~P.,  {Cooper} M.~C.,  {Pace} A.~B.,  {Boylan-Kolchin} M.,
  {Bullock} J.~S.,  {Garrison-Kimmel} S.,   {Wheeler} C.,  2016, \mn@doi
  [\mnras] {10.1093/mnras/stw2131}, \href
  {http://adsabs.harvard.edu/abs/2016MNRAS.463.1916F} {463, 1916}

\bibitem[\protect\citeauthoryear{{Fillingham}, {Cooper}, {Boylan-Kolchin},
  {Bullock}, {Garrison-Kimmel}  \& {Wheeler}}{{Fillingham}
  et~al.}{2018}]{Fillingham18}
{Fillingham} S.~P.,  {Cooper} M.~C.,  {Boylan-Kolchin} M.,  {Bullock} J.~S.,
  {Garrison-Kimmel} S.,   {Wheeler} C.,  2018, \mn@doi [\mnras]
  {10.1093/mnras/sty958}, \href
  {http://adsabs.harvard.edu/abs/2018MNRAS.477.4491F} {477, 4491}

\bibitem[\protect\citeauthoryear{{Fitts} et~al.,}{{Fitts}
  et~al.}{2017}]{Fitts17}
{Fitts} A.,  et~al., 2017, \mn@doi [\mnras] {10.1093/mnras/stx1757}, \href
  {http://adsabs.harvard.edu/abs/2017MNRAS.471.3547F} {471, 3547}

\bibitem[\protect\citeauthoryear{{Flores} \& {Primack}}{{Flores} \&
  {Primack}}{1994}]{Flores94}
{Flores} R.~A.,  {Primack} J.~R.,  1994, \mn@doi [\apjl] {10.1086/187350},
  \href {http://adsabs.harvard.edu/abs/1994ApJ...427L...1F} {427, L1}

\bibitem[\protect\citeauthoryear{{Gallart} et~al.,}{{Gallart}
  et~al.}{2015}]{Gallart15}
{Gallart} C.,  et~al., 2015, \mn@doi [\apjl] {10.1088/2041-8205/811/2/L18},
  \href {http://adsabs.harvard.edu/abs/2015ApJ...811L..18G} {811, L18}

\bibitem[\protect\citeauthoryear{{Garc{\'{\i}}a-Benito}
  et~al.,}{{Garc{\'{\i}}a-Benito} et~al.}{2017}]{GarciaBenito17}
{Garc{\'{\i}}a-Benito} R.,  et~al., 2017, \mn@doi [\aap]
  {10.1051/0004-6361/201731357}, \href
  {http://adsabs.harvard.edu/abs/2017A%26A...608A..27G} {608, A27}

\bibitem[\protect\citeauthoryear{{Garrison-Kimmel} et~al.,}{{Garrison-Kimmel}
  et~al.}{2017}]{SGK17}
{Garrison-Kimmel} S.,  et~al., 2017, preprint, \href
  {http://adsabs.harvard.edu/abs/2017arXiv170103792G} {} (\mn@eprint {arXiv}
  {1701.03792})

\bibitem[\protect\citeauthoryear{{Geha}, {Blanton}, {Yan}  \& {Tinker}}{{Geha}
  et~al.}{2012}]{Geha12}
{Geha} M.,  {Blanton} M.~R.,  {Yan} R.,   {Tinker} J.~L.,  2012, \mn@doi [\apj]
  {10.1088/0004-637X/757/1/85}, \href
  {http://adsabs.harvard.edu/abs/2012ApJ...757...85G} {757, 85}

\bibitem[\protect\citeauthoryear{{Gil de Paz} et~al.,}{{Gil de Paz}
  et~al.}{2005}]{Gil05}
{Gil de Paz} A.,  et~al., 2005, \mn@doi [\apjl] {10.1086/432054}, \href
  {http://adsabs.harvard.edu/abs/2005ApJ...627L..29G} {627, L29}

\bibitem[\protect\citeauthoryear{{Gil de Paz} et~al.,}{{Gil de Paz}
  et~al.}{2007}]{Gil07}
{Gil de Paz} A.,  et~al., 2007, \mn@doi [\apj] {10.1086/513730}, \href
  {http://adsabs.harvard.edu/abs/2007ApJ...661..115G} {661, 115}

\bibitem[\protect\citeauthoryear{{Graus}, {Bullock}, {Boylan-Kolchin}  \&
  {Weisz}}{{Graus} et~al.}{2016}]{Graus16}
{Graus} A.~S.,  {Bullock} J.~S.,  {Boylan-Kolchin} M.,   {Weisz} D.~R.,  2016,
  \mn@doi [\mnras] {10.1093/mnras/stv2728}, \href
  {http://adsabs.harvard.edu/abs/2016MNRAS.456..477G} {456, 477}

\bibitem[\protect\citeauthoryear{{Hahn} \& {Abel}}{{Hahn} \&
  {Abel}}{2011}]{Hahn11}
{Hahn} O.,  {Abel} T.,  2011, \mn@doi [\mnras]
  {10.1111/j.1365-2966.2011.18820.x}, \href
  {http://adsabs.harvard.edu/abs/2011MNRAS.415.2101H} {415, 2101}

\bibitem[\protect\citeauthoryear{{Harbeck} et~al.,}{{Harbeck}
  et~al.}{2001}]{Harbeck01}
{Harbeck} D.,  et~al., 2001, \mn@doi [\aj] {10.1086/324232}, \href
  {http://adsabs.harvard.edu/abs/2001AJ....122.3092H} {122, 3092}

\bibitem[\protect\citeauthoryear{{Hidalgo} et~al.,}{{Hidalgo}
  et~al.}{2013}]{Hidalgo13}
{Hidalgo} S.~L.,  et~al., 2013, \mn@doi [\apj] {10.1088/0004-637X/778/2/103},
  \href {http://adsabs.harvard.edu/abs/2013ApJ...778..103H} {778, 103}

\bibitem[\protect\citeauthoryear{{Hopkins}}{{Hopkins}}{2015}]{Hopkins15}
{Hopkins} P.~F.,  2015, \mn@doi [\mnras] {10.1093/mnras/stv195}, \href
  {http://adsabs.harvard.edu/abs/2015MNRAS.450...53H} {450, 53}

\bibitem[\protect\citeauthoryear{{Hopkins}}{{Hopkins}}{2017}]{Hopkins17diff}
{Hopkins} P.~F.,  2017, \mn@doi [\mnras] {10.1093/mnras/stw3306}, \href
  {http://adsabs.harvard.edu/abs/2017MNRAS.466.3387H} {466, 3387}

\bibitem[\protect\citeauthoryear{{Hopkins} et~al.,}{{Hopkins}
  et~al.}{2018}]{Hopkins18}
{Hopkins} P.~F.,  et~al., 2018, \mn@doi [\mnras] {10.1093/mnras/sty1690}, \href
  {http://adsabs.harvard.edu/abs/2018MNRAS.480..800H} {480, 800}

\bibitem[\protect\citeauthoryear{{Kacharov} et~al.,}{{Kacharov}
  et~al.}{2017}]{Kacharov17}
{Kacharov} N.,  et~al., 2017, \mn@doi [\mnras] {10.1093/mnras/stw3188}, \href
  {http://adsabs.harvard.edu/abs/2017MNRAS.466.2006K} {466, 2006}

\bibitem[\protect\citeauthoryear{{Katz} \& {White}}{{Katz} \&
  {White}}{1993}]{Katz93}
{Katz} N.,  {White} S.~D.~M.,  1993, \mn@doi [\apj] {10.1086/172935}, \href
  {http://adsabs.harvard.edu/abs/1993ApJ...412..455K} {412, 455}

\bibitem[\protect\citeauthoryear{{Knollmann} \& {Knebe}}{{Knollmann} \&
  {Knebe}}{2009}]{Knollmann09}
{Knollmann} S.~R.,  {Knebe} A.,  2009, \mn@doi [\apjs]
  {10.1088/0067-0049/182/2/608}, \href
  {http://adsabs.harvard.edu/abs/2009ApJS..182..608K} {182, 608}

\bibitem[\protect\citeauthoryear{{Kroupa}}{{Kroupa}}{2001}]{Kroupa01}
{Kroupa} P.,  2001, \mn@doi [\mnras] {10.1046/j.1365-8711.2001.04022.x}, \href
  {http://adsabs.harvard.edu/abs/2001MNRAS.322..231K} {322, 231}

\bibitem[\protect\citeauthoryear{{Krumholz} \& {Gnedin}}{{Krumholz} \&
  {Gnedin}}{2011}]{Krumholz11}
{Krumholz} M.~R.,  {Gnedin} N.~Y.,  2011, \mn@doi [\apj]
  {10.1088/0004-637X/729/1/36}, \href
  {http://adsabs.harvard.edu/abs/2011ApJ...729...36K} {729, 36}

\bibitem[\protect\citeauthoryear{{Larson}}{{Larson}}{1976}]{Larson76}
{Larson} R.~B.,  1976, \mn@doi [\mnras] {10.1093/mnras/176.1.31}, \href
  {http://adsabs.harvard.edu/abs/1976MNRAS.176...31L} {176, 31}

\bibitem[\protect\citeauthoryear{{Leaman} et~al.,}{{Leaman}
  et~al.}{2017}]{Leaman17}
{Leaman} R.,  et~al., 2017, \mn@doi [\mnras] {10.1093/mnras/stx2014}, \href
  {http://adsabs.harvard.edu/abs/2017MNRAS.472.1879L} {472, 1879}

\bibitem[\protect\citeauthoryear{{Leitner}}{{Leitner}}{2012}]{Leitner12}
{Leitner} S.~N.,  2012, \mn@doi [\apj] {10.1088/0004-637X/745/2/149}, \href
  {https://ui.adsabs.harvard.edu/\#abs/2012ApJ...745..149L} {745, 149}

\bibitem[\protect\citeauthoryear{{Makarova}, {Makarov}, {Karachentsev}, {Tully}
   \& {Rizzi}}{{Makarova} et~al.}{2017}]{Makarova17}
{Makarova} L.~N.,  {Makarov} D.~I.,  {Karachentsev} I.~D.,  {Tully} R.~B.,
  {Rizzi} L.,  2017, \mn@doi [\mnras] {10.1093/mnras/stw2502}, \href
  {http://adsabs.harvard.edu/abs/2017MNRAS.464.2281M} {464, 2281}

\bibitem[\protect\citeauthoryear{{Martin} et~al.,}{{Martin}
  et~al.}{2005}]{Martin05}
{Martin} D.~C.,  et~al., 2005, \mn@doi [\apjl] {10.1086/426387}, \href
  {http://adsabs.harvard.edu/abs/2005ApJ...619L...1M} {619, L1}

\bibitem[\protect\citeauthoryear{{Mart{\'{\i}}nez-V{\'a}zquez}
  et~al.,}{{Mart{\'{\i}}nez-V{\'a}zquez} et~al.}{2015}]{Martinez15}
{Mart{\'{\i}}nez-V{\'a}zquez} C.~E.,  et~al., 2015, \mn@doi [\mnras]
  {10.1093/mnras/stv2014}, \href
  {http://adsabs.harvard.edu/abs/2015MNRAS.454.1509M} {454, 1509}

\bibitem[\protect\citeauthoryear{{McConnachie}}{{McConnachie}}{2012}]{McConnachie12}
{McConnachie} A.~W.,  2012, \mn@doi [\aj] {10.1088/0004-6256/144/1/4}, \href
  {http://adsabs.harvard.edu/abs/2012AJ....144....4M} {144, 4}

\bibitem[\protect\citeauthoryear{{McMonigal} et~al.,}{{McMonigal}
  et~al.}{2014}]{McMonigal14}
{McMonigal} B.,  et~al., 2014, \mn@doi [\mnras] {10.1093/mnras/stu1659}, \href
  {http://adsabs.harvard.edu/abs/2014MNRAS.444.3139M} {444, 3139}

\bibitem[\protect\citeauthoryear{{McQuinn} et~al.,}{{McQuinn}
  et~al.}{2017}]{McQuinn17}
{McQuinn} K.~B.~W.,  et~al., 2017, \mn@doi [\apj] {10.3847/1538-4357/834/1/78},
  \href {http://adsabs.harvard.edu/abs/2017ApJ...834...78M} {834, 78}

\bibitem[\protect\citeauthoryear{{Mo}, {Mao}  \& {White}}{{Mo}
  et~al.}{1998}]{Mo98}
{Mo} H.~J.,  {Mao} S.,   {White} S.~D.~M.,  1998, \mn@doi [\mnras]
  {10.1046/j.1365-8711.1998.01227.x}, \href
  {http://adsabs.harvard.edu/abs/1998MNRAS.295..319M} {295, 319}

\bibitem[\protect\citeauthoryear{{Monelli} et~al.,}{{Monelli}
  et~al.}{2010}]{Monelli10}
{Monelli} M.,  et~al., 2010, \mn@doi [\apj] {10.1088/0004-637X/722/2/1864},
  \href {http://adsabs.harvard.edu/abs/2010ApJ...722.1864M} {722, 1864}

\bibitem[\protect\citeauthoryear{{Monelli} et~al.,}{{Monelli}
  et~al.}{2016}]{Monelli16}
{Monelli} M.,  et~al., 2016, \mn@doi [\apj] {10.3847/0004-637X/819/2/147},
  \href {http://adsabs.harvard.edu/abs/2016ApJ...819..147M} {819, 147}

\bibitem[\protect\citeauthoryear{{Moore}}{{Moore}}{1994}]{Moore94}
{Moore} B.,  1994, \mn@doi [\nat] {10.1038/370629a0}, \href
  {http://adsabs.harvard.edu/abs/1994Natur.370..629M} {370, 629}

\bibitem[\protect\citeauthoryear{{Nelson} et~al.,}{{Nelson}
  et~al.}{2012}]{Nelson12}
{Nelson} E.~J.,  et~al., 2012, \mn@doi [\apjl] {10.1088/2041-8205/747/2/L28},
  \href {http://adsabs.harvard.edu/abs/2012ApJ...747L..28N} {747, L28}

\bibitem[\protect\citeauthoryear{{Nelson} et~al.,}{{Nelson}
  et~al.}{2016}]{Nelson16}
{Nelson} E.~J.,  et~al., 2016, \mn@doi [\apj] {10.3847/0004-637X/828/1/27},
  \href {http://adsabs.harvard.edu/abs/2016ApJ...828...27N} {828, 27}

\bibitem[\protect\citeauthoryear{{O{\~n}orbe}, {Garrison-Kimmel}, {Maller},
  {Bullock}, {Rocha}  \& {Hahn}}{{O{\~n}orbe} et~al.}{2014}]{Onorbe14}
{O{\~n}orbe} J.,  {Garrison-Kimmel} S.,  {Maller} A.~H.,  {Bullock} J.~S.,
  {Rocha} M.,   {Hahn} O.,  2014, \mn@doi [\mnras] {10.1093/mnras/stt2020},
  \href {http://adsabs.harvard.edu/abs/2014MNRAS.437.1894O} {437, 1894}

\bibitem[\protect\citeauthoryear{{O{\~n}orbe}, {Boylan-Kolchin}, {Bullock},
  {Hopkins}, {Kere{\v s}}, {Faucher-Gigu{\`e}re}, {Quataert}  \&
  {Murray}}{{O{\~n}orbe} et~al.}{2015}]{Onorbe15}
{O{\~n}orbe} J.,  {Boylan-Kolchin} M.,  {Bullock} J.~S.,  {Hopkins} P.~F.,
  {Kere{\v s}} D.,  {Faucher-Gigu{\`e}re} C.-A.,  {Quataert} E.,   {Murray} N.,
   2015, \mn@doi [\mnras] {10.1093/mnras/stv2072}, \href
  {http://adsabs.harvard.edu/abs/2015MNRAS.454.2092O} {454, 2092}

\bibitem[\protect\citeauthoryear{{Okamoto}, {Arimoto}, {Tolstoy}, {Jablonka},
  {Irwin}, {Komiyama}, {Yamada}  \& {Onodera}}{{Okamoto}
  et~al.}{2017}]{Okamoto17}
{Okamoto} S.,  {Arimoto} N.,  {Tolstoy} E.,  {Jablonka} P.,  {Irwin} M.~J.,
  {Komiyama} Y.,  {Yamada} Y.,   {Onodera} M.,  2017, \mn@doi [\mnras]
  {10.1093/mnras/stx086}, \href
  {http://adsabs.harvard.edu/abs/2017MNRAS.467..208O} {467, 208}

\bibitem[\protect\citeauthoryear{{Patel} et~al.,}{{Patel}
  et~al.}{2013}]{Patel13}
{Patel} S.~G.,  et~al., 2013, \mn@doi [\apj] {10.1088/0004-637X/766/1/15},
  \href {http://adsabs.harvard.edu/abs/2013ApJ...766...15P} {766, 15}

\bibitem[\protect\citeauthoryear{{Pilkington} et~al.,}{{Pilkington}
  et~al.}{2012}]{Pilkington12}
{Pilkington} K.,  et~al., 2012, \mn@doi [\aap] {10.1051/0004-6361/201117466},
  \href {http://adsabs.harvard.edu/abs/2012A%26A...540A..56P} {540, A56}

\bibitem[\protect\citeauthoryear{{Pontzen} \& {Governato}}{{Pontzen} \&
  {Governato}}{2012}]{Pontzen12}
{Pontzen} A.,  {Governato} F.,  2012, \mn@doi [\mnras]
  {10.1111/j.1365-2966.2012.20571.x}, \href
  {http://adsabs.harvard.edu/abs/2012MNRAS.421.3464P} {421, 3464}

\bibitem[\protect\citeauthoryear{{Read}, {Agertz}  \& {Collins}}{{Read}
  et~al.}{2016}]{Read16}
{Read} J.~I.,  {Agertz} O.,   {Collins} M.~L.~M.,  2016, \mn@doi [\mnras]
  {10.1093/mnras/stw713}, \href
  {http://adsabs.harvard.edu/abs/2016MNRAS.459.2573R} {459, 2573}

\bibitem[\protect\citeauthoryear{{Ricotti} \& {Gnedin}}{{Ricotti} \&
  {Gnedin}}{2005}]{Ricotti05}
{Ricotti} M.,  {Gnedin} N.~Y.,  2005, \mn@doi [\apj] {10.1086/431415}, \href
  {http://adsabs.harvard.edu/abs/2005ApJ...629..259R} {629, 259}

\bibitem[\protect\citeauthoryear{{Rodriguez Wimberly}, {Cooper}, {Fillingham},
  {Boylan-Kolchin}, {Bullock}  \& {Garrison-Kimmel}}{{Rodriguez Wimberly}
  et~al.}{2018}]{Wimberly18}
{Rodriguez Wimberly} M.~K.,  {Cooper} M.~C.,  {Fillingham} S.~P.,
  {Boylan-Kolchin} M.,  {Bullock} J.~S.,   {Garrison-Kimmel} S.,  2018,
  preprint, \href {http://adsabs.harvard.edu/abs/2018arXiv180607891R} {}
  (\mn@eprint {arXiv} {1806.07891})

\bibitem[\protect\citeauthoryear{{Santana} et~al.,}{{Santana}
  et~al.}{2016}]{Santana16}
{Santana} F.~A.,  et~al., 2016, \mn@doi [\apj] {10.3847/0004-637X/829/2/86},
  \href {http://adsabs.harvard.edu/abs/2016ApJ...829...86S} {829, 86}

\bibitem[\protect\citeauthoryear{{Sawala} et~al.,}{{Sawala}
  et~al.}{2016}]{Sawala16b}
{Sawala} T.,  et~al., 2016, \mn@doi [\mnras] {10.1093/mnras/stw145}, \href
  {http://adsabs.harvard.edu/abs/2016MNRAS.457.1931S} {457, 1931}

\bibitem[\protect\citeauthoryear{{Skillman} et~al.,}{{Skillman}
  et~al.}{2017}]{Skillman17}
{Skillman} E.~D.,  et~al., 2017, \mn@doi [\apj] {10.3847/1538-4357/aa60c5},
  \href {http://adsabs.harvard.edu/abs/2017ApJ...837..102S} {837, 102}

\bibitem[\protect\citeauthoryear{{Starkenburg} \& {Helmi}}{{Starkenburg} \&
  {Helmi}}{2015}]{Starkenburg15}
{Starkenburg} T.~K.,  {Helmi} A.,  2015, \mn@doi [\aap]
  {10.1051/0004-6361/201425082}, \href
  {http://adsabs.harvard.edu/abs/2015A%26A...575A..59S} {575, A59}

\bibitem[\protect\citeauthoryear{{Stinson}, {Dalcanton}, {Quinn}, {Gogarten},
  {Kaufmann}  \& {Wadsley}}{{Stinson} et~al.}{2009}]{Stinson09}
{Stinson} G.~S.,  {Dalcanton} J.~J.,  {Quinn} T.,  {Gogarten} S.~M.,
  {Kaufmann} T.,   {Wadsley} J.,  2009, \mn@doi [\mnras]
  {10.1111/j.1365-2966.2009.14555.x}, \href
  {http://adsabs.harvard.edu/abs/2009MNRAS.395.1455S} {395, 1455}

\bibitem[\protect\citeauthoryear{{Su}, {Hopkins}, {Hayward},
  {Faucher-Gigu{\`e}re}, {Kere{\v s}}, {Ma}  \& {Robles}}{{Su}
  et~al.}{2017}]{Su17}
{Su} K.-Y.,  {Hopkins} P.~F.,  {Hayward} C.~C.,  {Faucher-Gigu{\`e}re} C.-A.,
  {Kere{\v s}} D.,  {Ma} X.,   {Robles} V.~H.,  2017, \mn@doi [\mnras]
  {10.1093/mnras/stx1463}, \href
  {http://adsabs.harvard.edu/abs/2017MNRAS.471..144S} {471, 144}

\bibitem[\protect\citeauthoryear{{Tolstoy}, {Hill}  \& {Tosi}}{{Tolstoy}
  et~al.}{2009}]{Tolstoy09}
{Tolstoy} E.,  {Hill} V.,   {Tosi} M.,  2009, \mn@doi [\araa]
  {10.1146/annurev-astro-082708-101650}, \href
  {http://adsabs.harvard.edu/abs/2009ARA%26A..47..371T} {47, 371}

\bibitem[\protect\citeauthoryear{{Walmswell}, {Eldridge}, {Brewer}  \&
  {Tout}}{{Walmswell} et~al.}{2013}]{Walmswell13}
{Walmswell} J.~J.,  {Eldridge} J.~J.,  {Brewer} B.~J.,   {Tout} C.~A.,  2013,
  \mn@doi [\mnras] {10.1093/mnras/stt1444}, \href
  {http://adsabs.harvard.edu/abs/2013MNRAS.435.2171W} {435, 2171}

\bibitem[\protect\citeauthoryear{{Weisz}, {Dolphin}, {Skillman}, {Holtzman},
  {Gilbert}, {Dalcanton}  \& {Williams}}{{Weisz} et~al.}{2014a}]{Weisz14}
{Weisz} D.~R.,  {Dolphin} A.~E.,  {Skillman} E.~D.,  {Holtzman} J.,  {Gilbert}
  K.~M.,  {Dalcanton} J.~J.,   {Williams} B.~F.,  2014a, \mn@doi [\apj]
  {10.1088/0004-637X/789/2/147}, \href
  {http://adsabs.harvard.edu/abs/2014ApJ...789..147W} {789, 147}

\bibitem[\protect\citeauthoryear{{Weisz}, {Johnson}  \& {Conroy}}{{Weisz}
  et~al.}{2014b}]{Weisz14b}
{Weisz} D.~R.,  {Johnson} B.~D.,   {Conroy} C.,  2014b, \mn@doi [\apjl]
  {10.1088/2041-8205/794/1/L3}, \href
  {http://adsabs.harvard.edu/abs/2014ApJ...794L...3W} {794, L3}

\bibitem[\protect\citeauthoryear{{Weisz}, {Dolphin}, {Skillman}, {Holtzman},
  {Gilbert}, {Dalcanton}  \& {Williams}}{{Weisz} et~al.}{2015}]{Weisz15}
{Weisz} D.~R.,  {Dolphin} A.~E.,  {Skillman} E.~D.,  {Holtzman} J.,  {Gilbert}
  K.~M.,  {Dalcanton} J.~J.,   {Williams} B.~F.,  2015, \mn@doi [\apj]
  {10.1088/0004-637X/804/2/136}, \href
  {http://adsabs.harvard.edu/abs/2015ApJ...804..136W} {804, 136}

\bibitem[\protect\citeauthoryear{{Wetzel}, {Tollerud}  \& {Weisz}}{{Wetzel}
  et~al.}{2015}]{Wetzel15}
{Wetzel} A.~R.,  {Tollerud} E.~J.,   {Weisz} D.~R.,  2015, \mn@doi [\apjl]
  {10.1088/2041-8205/808/1/L27}, \href
  {http://adsabs.harvard.edu/abs/2015ApJ...808L..27W} {808, L27}

\bibitem[\protect\citeauthoryear{{del Pino}, {Aparicio}  \& {Hidalgo}}{{del
  Pino} et~al.}{2015}]{delPino15}
{del Pino} A.,  {Aparicio} A.,   {Hidalgo} S.~L.,  2015, \mn@doi [\mnras]
  {10.1093/mnras/stv2174}, \href
  {http://adsabs.harvard.edu/abs/2015MNRAS.454.3996D} {454, 3996}

\bibitem[\protect\citeauthoryear{{van Dokkum} et~al.,}{{van Dokkum}
  et~al.}{2010}]{vanDokkum10}
{van Dokkum} P.~G.,  et~al., 2010, \mn@doi [\apj]
  {10.1088/0004-637X/709/2/1018}, \href
  {http://adsabs.harvard.edu/abs/2010ApJ...709.1018V} {709, 1018}

\makeatother
\end{thebibliography}




\appendix

\begin{figure*}
	\includegraphics[width=0.4\textwidth, trim = 0 0 0 0]{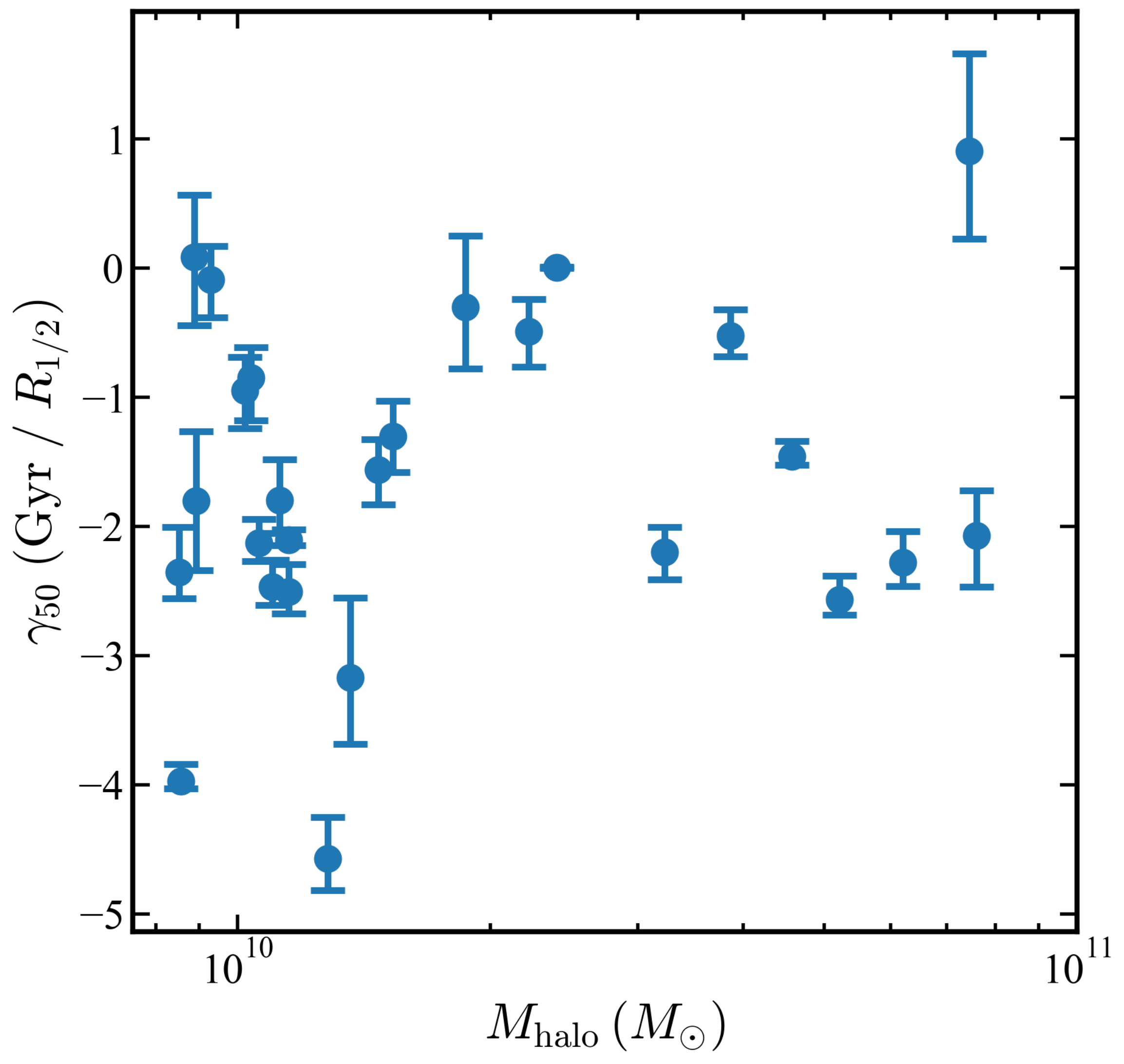}
	\includegraphics[width=0.4\textwidth, trim = 0 0 0 0]{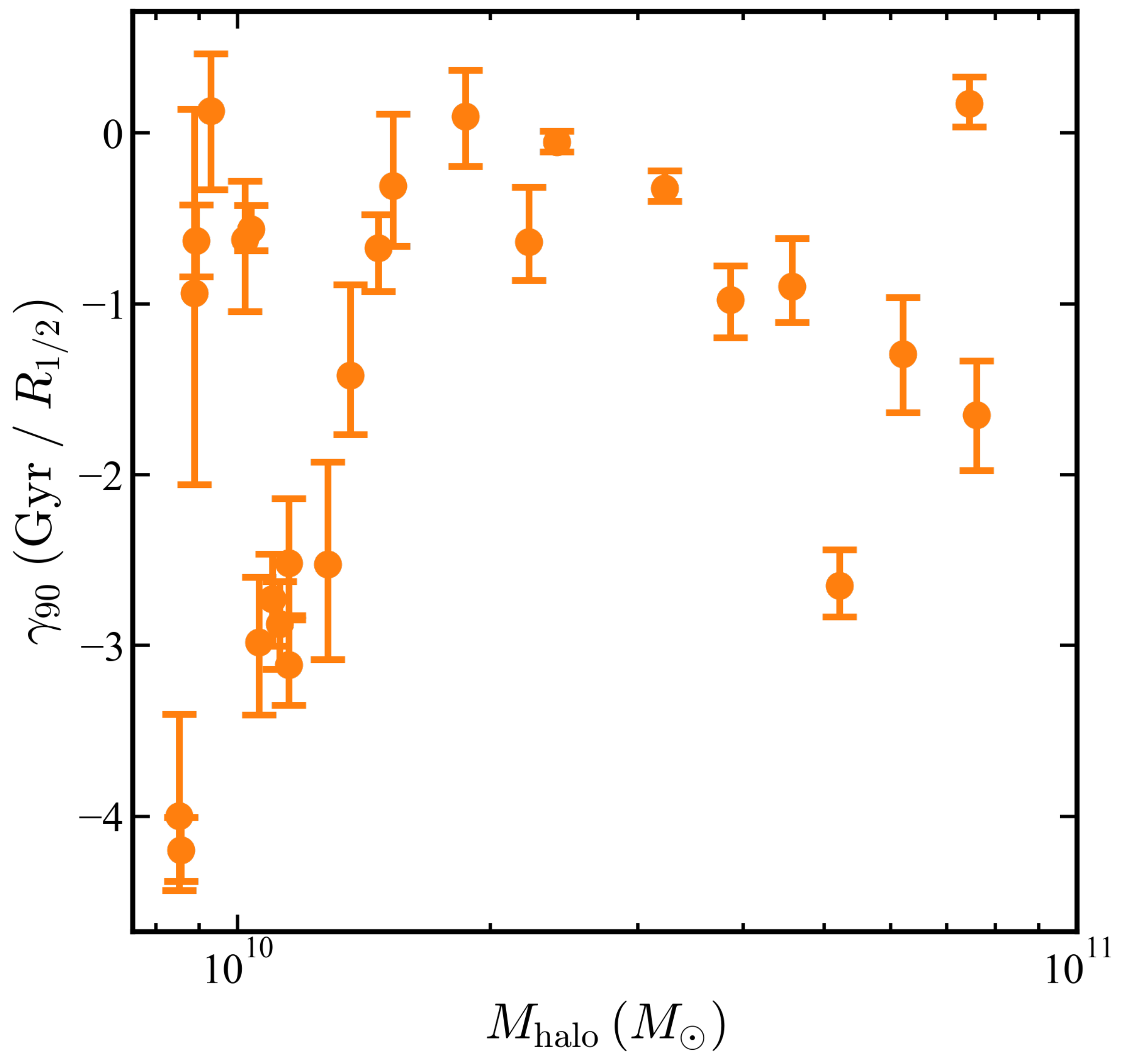}
    \includegraphics[width=0.4\textwidth, trim = 0 0 0 0]{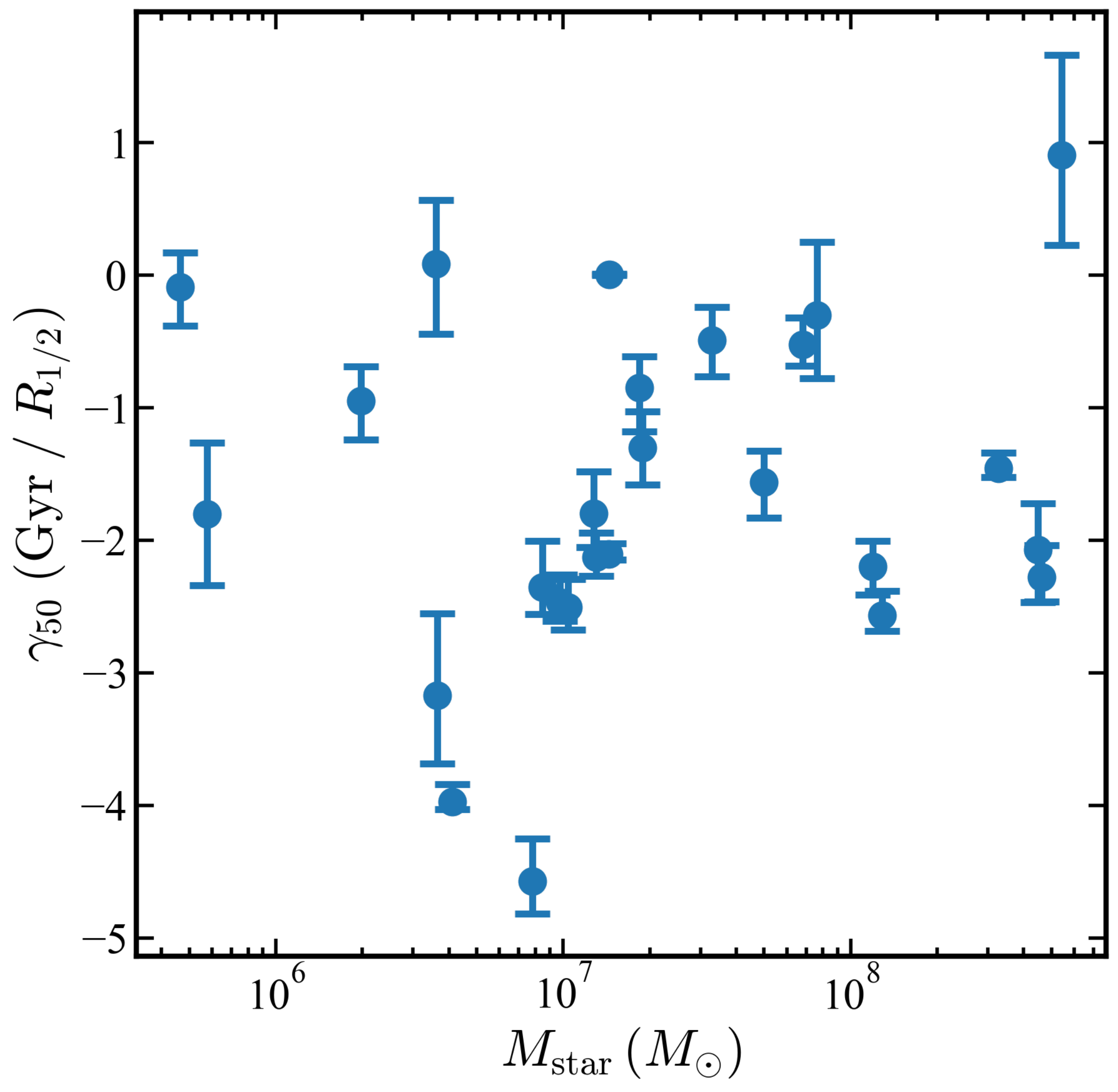}
	\includegraphics[width=0.4\textwidth, trim = 0 0 0 0]{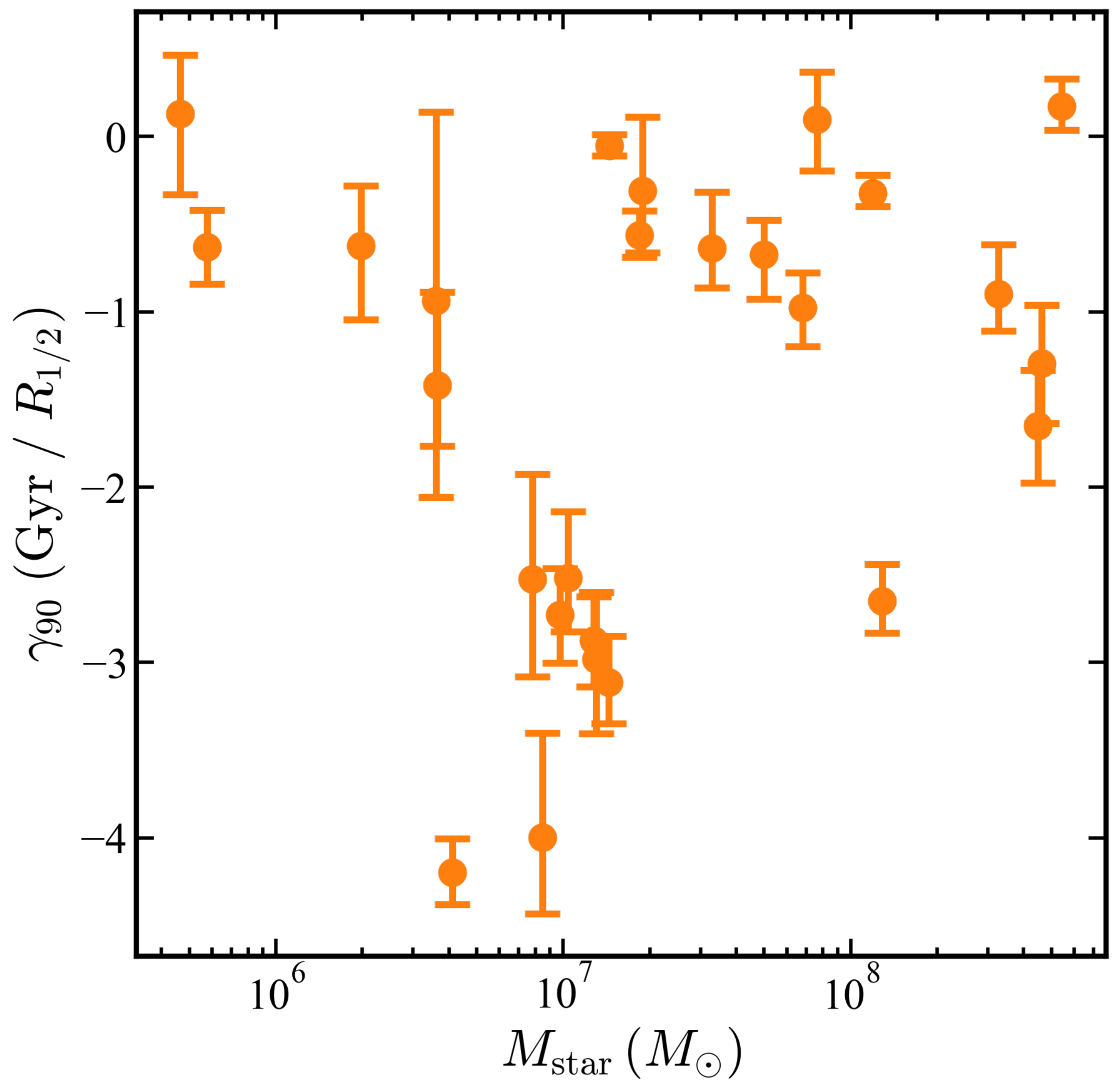}
    \includegraphics[width=0.4\textwidth, trim = 0 0 0 0]{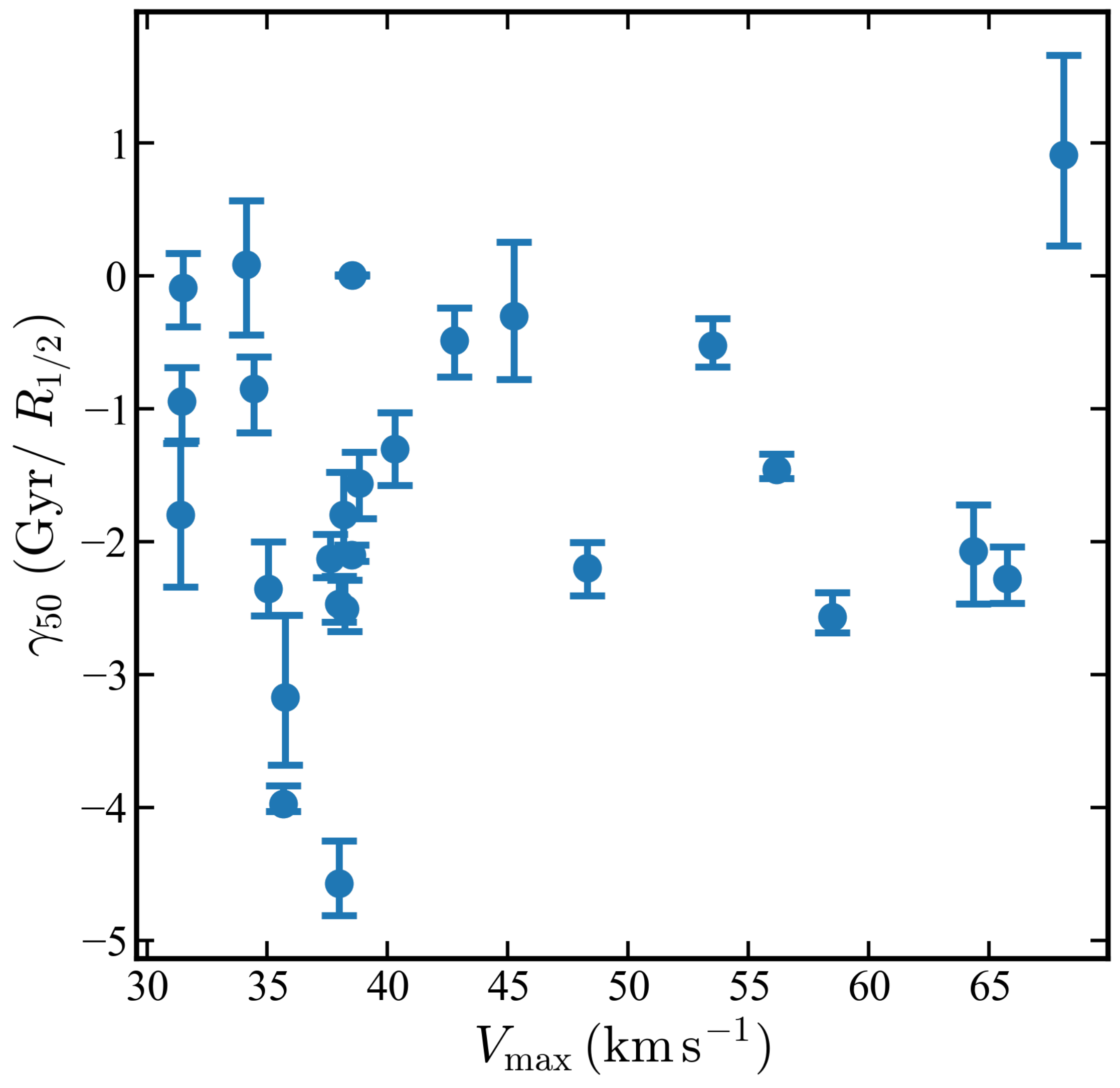}
	\includegraphics[width=0.4\textwidth, trim = 0 0 0 0]{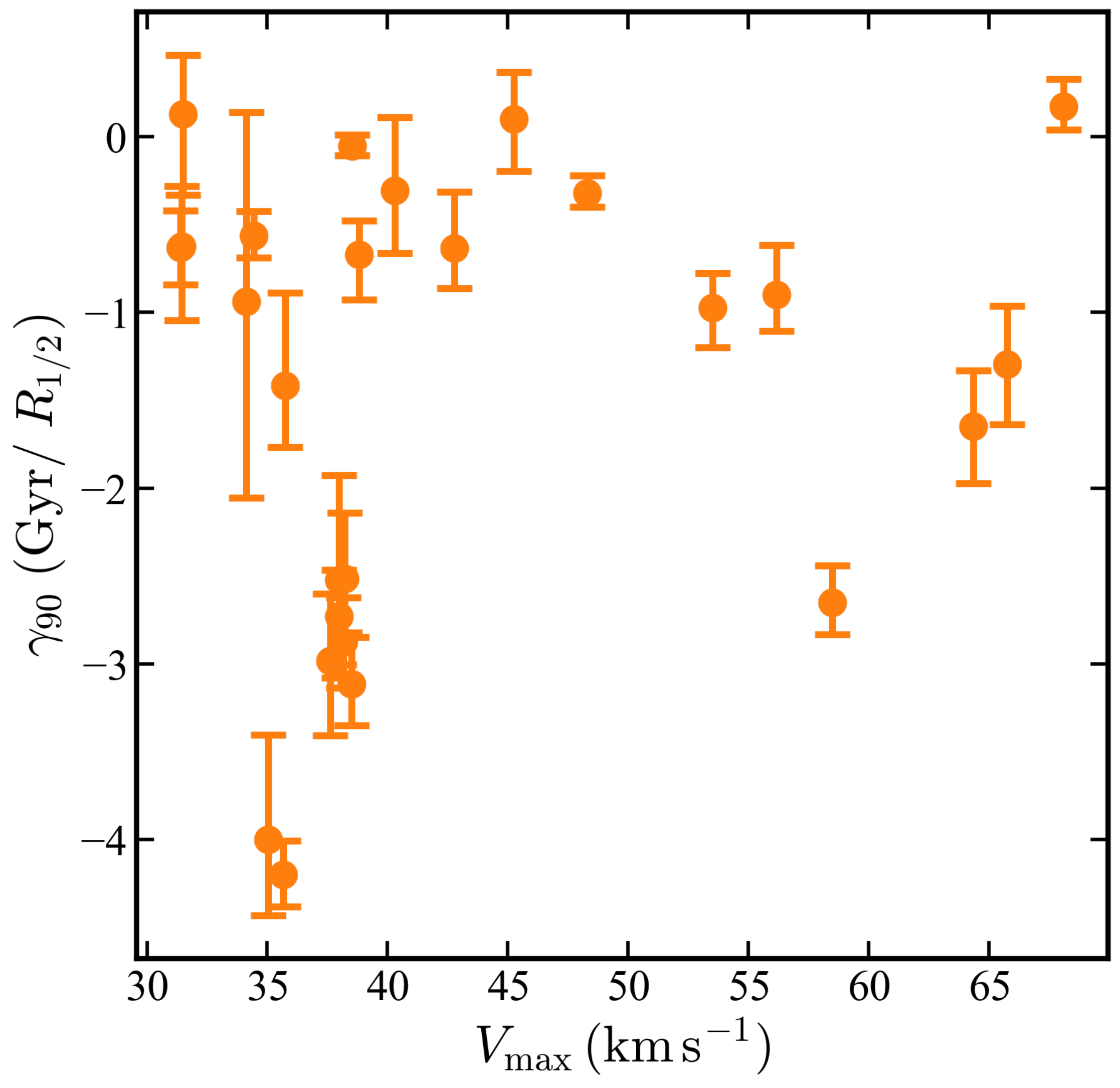}
	\centering
	\caption[]{Age gradients for all of the galaxies in the sample, plotted against $M_{\rm vir}$ and $M_{\rm star}$.  No trend is apparent. Note that the galaxies in the \cite{Fitts17} sample were specifically selected to lie at $M_{\rm halo}$ = $10^{10}$ M$_{\odot}$.}
	\label{fig:slope_relations}
\end{figure*}

\section{Age Gradient vs. Mass}

Figure \ref{fig:slope_relations} shows the relationship between age gradients and halo mass and stellar mass.  Unlike the clear trends seen in Figure \ref{fig:slope_time_relation} between gradient and age, there appears to be no correlation with these other common parameters.


\bsp	
\label{lastpage}
\end{document}